\documentclass[pdflatex,sn-mathphys,Numbered]{sn-jnl}

\usepackage{graphicx}%
\usepackage{multirow}%
\usepackage{amsmath,amssymb,amsfonts}%
\usepackage{amsthm}%
\usepackage{mathrsfs}%
\usepackage[title]{appendix}%
\usepackage{xcolor}%
\usepackage{textcomp}%
\usepackage{manyfoot}%
\usepackage{booktabs}%
\usepackage{algorithmicx}%
\usepackage{algpseudocode}%
\usepackage{listings}%
\usepackage{natbib}

\usepackage{graphicx}
\usepackage{amsfonts}
\usepackage{multirow}
\usepackage{cases}
\usepackage{booktabs}
\usepackage{multicol}
\usepackage{booktabs}
\usepackage{tabularx}
\usepackage{anyfontsize}

\usepackage{algorithm2e,epsfig,amssymb,amsmath,natbib,listings,lineno,color,cleveref}
\usepackage{subfig,float}
\usepackage{color}

\definecolor{orange}{rgb}{1,0.5,0}
\definecolor{amethyst}{rgb}{0.6,0.4,0.8}
\definecolor{aureolin}{rgb}{0.99,0.93,0.0}
\definecolor{awesome}{rgb}{1.0,0.13,0.32}
\definecolor{ao-green}{rgb}{0.0, 0.5, 0.0}

\newcommand{\review}[1]{\textcolor{black}{#1}}

\begin{document}

\title[Article Title]{Non-dissipative large-eddy simulation of wall-bounded transcritical turbulent flows}


\author*[1]{\fnm{Marc} \sur{Bernades}}\email{marc.bernades@upc.edu}

\author*[2]{\fnm{Florent} \sur{Duchaine}}\email{florent.duchaine@cerfacs.fr}

\author*[1]{\fnm{Francesco} \sur{Capuano}}\email{francesco.capuano@upc.edu}

\author*[1]{\fnm{Llu\'is} \sur{Jofre}}\email{lluis.jofre@upc.edu}

\affil*[1]{\orgdiv{Department of Fluid Mechanics}, \orgname{Universitat Polit\`ecnica de Catalunya $\cdot$ BarcelonaTech (UPC)}, \orgaddress{\city{Barcelona}, \postcode{08034}, \country{Spain}}}

\affil*[2]{\orgname{CFD Team, CERFACS}, \orgaddress{\city{Toulouse}, \postcode{31057}, \country{France}}}



\abstract{
\textit{A posteriori} analysis based upon a recently proposed non-dissipative large-eddy simulation framework for transcritical wall-bounded turbulence has been carried out.
Due to the complexities arisen in such flows, the discretization requires kinetic-energy- and pressure-equilibrium-preservation schemes to yield stable and non-dissipative scale-resolving simulations.
On the basis of this framework, the objectives are to (i) compute wall-resolved and wall-modeled large-eddy simulations of a high-pressure transcritical turbulent channel flow, and (ii) assess the thermofluid performance with respect to a direct numerical simulation.
In this regard, three different subgrid-scale stress tensor models have been considered, together with models for the unresolved scales of the filtered pressure transport equation and equation of state.
In terms of wall-modeling, models based on the ``standard law of the wall'' and velocity-temperature coupled approaches have been assessed.
The results show that for the wall-resolved approaches, the subgrid-scale stress tensors examined slightly deviate from the time-averaged velocity and temperature reference profiles.
In terms of bulk performance, it has been found that the Nusselt number and skin-friction coefficient are relatively well captured at the cold and hot walls, respectively. While heat transfer phenomena are fairly well reproduced, particularly the Prandtl and heat flux trends along wall-normal direction.
Additionally, the wall-modeled strategies improve the recovery of first-order statistics, although they do not attain the profile in the log-law region. However, they enhance the wall metrics and the heat flux prediction.
It is, thus, concluded that dedicated efforts by the research community are needed to improve the prediction accuracy of existing subgrid-scale and wall models for wall-bounded transcritical turbulence.
}

\keywords{Large-eddy simulation, Turbulence modeling, Supercritical fluids}



\maketitle

\section{Introduction}\label{sec1}

Wall-bounded high-pressure transcritical turbulent flows are relevant in many engineering applications, like for example energy conversion and propulsion systems.
They exhibit two main advantages with respect to atmospheric confined flows: (i) superior heat-transfer performance and mixing rates~\cite{Bernades2022-A,Bernades2023b-A}, and (ii) the fluid properties in the transcritical regime can be leveraged to maximize turbulent intensity to, for example, develop microconfined turbulence~\citep{Bernades2022-A,Jofre2023-A}.
Nevertheless, as a result of the significant thermophysical variations~\citep{Jofre2020-A,Jofre2021-A,Bernades2023a-A,Masclans2023-A}, computational research efforts are still necessary to accurately simulate the thermo-fluid behavior of such type of flows when they operate across the pseudo-boiling line (above the critical point).
In particular, at supercritical fluid conditions, intermolecular forces and finite packing volume effects become important presenting two thermodynamic states: supercritical gas-like and supercritical liquid-like~\cite{Jofre2020-A,Jofre2021-A}.
Therefore, high-fidelity computational simulations become an essential tool to elucidate the underlying physics of high-pressure transcritical fluids turbulence by means of either direct numerical simulations (DNS) and/or large-eddy simulations (LES).
DNS provides higher levels of accuracy as the meshes utilized are fine enough to capture the smallest eddies (Kolmogorov scale), but at the expense of large computational resources.
Instead, LES approaches are favored in high-Reynolds-number engineering applications to reduce such spatio-temporal requirements inherent in DNS and mitigate the computational complexity of the problem.

LES is an efficient approach to study the physics of turbulence based on filtering out the small scales of the flow.
Nonetheless, the small-scale motions and their effects on the resolved flow field are not negligible, and therefore require supplementary modeling~\citep{Sagaut2009-B,Jofre2018-A,Jofre2019-A}, the so-called subgrid scale (SGS) models.
The development and assessment of such models for \textit{closing} the resulting SGS terms is a very active field of research in a wide range of multiphysics turbulent flow regimes and applications~\cite{Pitsch2006-A,Georgiadis2010-A,Fox2012-A}.
However, LES modeling for trans/supercritical thermodynamic conditions is still in its infancy as it requires the modeling of additional SGS terms not present in low-pressure modeling frameworks, but that may become important at high pressures~\cite{Selle2007-A, Borghesi2015-A}.
The unavailability of such SGS models for high-pressure transcritical flows, forced~\citet{Schmitt2009-A,Ren2022-A,Wang2022-A}, among others, to use classical closure models, such as Smagorinsky~\cite{Smagorinsky1963-A} and wall-adapting local eddy-viscosity (WALE)~\cite{Nicoud1999-A,Ducros1999-A}.
In fact,~\citet{Muller2016-A} assessed SGS models for cyrogenic injection at supercritical pressures and reported that the relative influence of physical and numerical model uncertainties strongly depends on the thermodynamic regime.
Thus, with the objective of assessing current existing SGS models, \citet{Giauque2024-A} performed \textit{a priori} analysis and proposed a SGS stress tensor model for homogeneous isotropic turbulence (HIT) under real-gas conditions.
More recently, research work by~\citet{Bernades2024-A} has aimed to propose a novel LES framework for high-pressure transcritical fluid turbulence based on a novel numerical scheme which attains kinetic-energy- and pressure-equilibrium-preservation~\cite{Bernades2023b-A}.
In detail, different SGS stress tensor models (eddy-viscosity- and scale-similarity-like) were evaluated reporting relatively poor performance compared against filtered DNS data.
In addition, the assessment of the (i) relative importance of each of the different unclosed SGS terms identified that some of them can be safely neglected, and (ii) closure expressions for the unresolved terms of the pressure transport equation and the nonlinear equation of state were proposed.

LES has been employed in numerous types of turbulent flows since its first developments~\cite{Smagorinsky1963-A}, ranging from canonical turbulent channel flows and mixing layers~\cite{Moin1998-A,Piomelli1988-A} to complex industrial combustion and turbomachinery flows~\cite{DiMare2004-A,Giauque2005-A}.
Nevertheless, high-pressure transcritical LES frameworks are based on SGS stress tensor models which are not specifically developed for such flows, viz. classical SGS models developed for incompressible and/or low-pressure applications such as Smagorinsky, WALE, Sigma~\cite{Toda2010-A} and scale-similarity~\cite{Vreman1995-B}.
Moreover, as introduced above, additional SGS terms appear when dealing with the thermophysical nonlinearities present at high-pressure transcritical fluid regimes, which have been traditionally omitted in previous works.
Furthermore, it is well-known that flow scales become exponentially smaller close to walls.
Therefore, the mesh resolution requirements to capture the boundary layers rapidly become exceedingly unfeasible when the Reynolds number increases.
This problematic has motivated the development of wall models~\cite{Bose2018-A} to alleviate the mesh resolution requirements at walls by means of additional modeling and, consequently, depending whether the near-wall eddies are resolved or modeled LES approaches are termed either wall-resolved (WRLES) or wall-modeled (WMLES), respectively.
Although the former has a more favorable scaling than DNS, it is still inappropriate for realistic engineering configurations. Instead, the latter requires additional boundary conditions to account for the unresolved part of the boundary layer coupled with the outer-wall flow. 
Consequently, WMLES alleviates the cost induced in the near-wall regions by combining SGS with wall-stress models, such as the classic/standard log-law~\cite{VanDriest1951-A} or more recent methods~\cite{Larsson2015-A,Bose2018-A}.
However, previous works have shown that non-isothermal flows with variable density and viscosity~\cite{Bernades2023a-A} do not collapse to the ``standard law of the wall'', even when utilizing alternative transformations specifically suited for non-isothermal compressible boundary layers~\cite{Trettel2016-A}.
In the case of wall-bounded high-pressure transcritical turbulent flows, this mismatch becomes important in the log-law region and is caused by the presence of a baroclinic torque in the pseudo-boiling region.
In this regard,~\citet{Cabrit2009-B} proposed a \textit{coupled} wall-model to account for the effects of the molecular Prandtl number and strong temperature/density gradients, which is consistent with classical logarithmic formulations for velocity and temperature.
This WMLES was assessed for isothermal low-Mach-number turbulent channel flows to predict the wall shear stress and heat fluxes.
It was found that the coupling between momentum and energy improved the accuracy by correctly recovering the velocity and temperature profiles~\cite{Cabrit2009-A}.
In addition, \citet{Potier2018-B} utilized this coupled model for turbulent channel flows at high temperature ratios in conjunction with the Smagorinsky model and reported similar improvements, but without properly estimating the Nusselt number.
It was particularly found that the Nusselt number was strongly sensitive to the SGS stress model utilized.
Therefore, the \textit{a posteriori} analysis performed in this work aims to assess three SGS stress tensor models typically utilized in turbulent flow engineering applications~\cite{Vadrot2021-B} (Smagorinsky, Dynamic Smagorinsky and WALE) to a canonical non-isothermal high-pressure transcritical turbulent channel flow based on both WRLES and WMLES (standard and coupled) strategies following the insights of the \textit{a priori} analysis performed by~\citet{Bernades2024-A}.
The results are benchmarked against an equivalent fully-resolved reference DNS dataset~\cite{Bernades2023a-A}.

The paper is organized as follows.
First, in Section~\ref{sec:flow_physics_modeling}, the flow physics modeling, real-gas equation of state and LES SGS closure expressions are presented.
This is followed by Section~\ref{sec:computational_approaches}, which introduces the WRLES and WMLES approaches utilized.
Next, the computational setup of the problem is described in Section~\ref{sec:results_LES}, together with the mesh resolutions employed and the results of first-order flow statistics.
The performance of the models in terms of heat transfer phenomena is subsequently analyzed in Section~\ref{sec:heat_transfer}.
Finally, Section~\ref{sec:conclusions} reports concluding remarks and proposes future research directions.

\section{Flow physics modeling}  \label{sec:flow_physics_modeling}

The LES framework for the turbulent flow motion of supercritical fluids is described by the following set of transport equations of mass, momentum and pressure
\begin{align}
& \frac{\partial \rho}{\partial t} + \nabla \cdot \left( \rho \mathbf{u} \right) = 0,    \label{eq:mass} \\
& \frac{\partial \left( \rho \mathbf{u} \right) }{\partial t} + \nabla \cdot \left( \rho \mathbf{u} \mathbf{u} \right) = -\nabla P +\nabla \cdot \boldsymbol{\sigma} - \boldsymbol{\alpha}_1,  \label{eq:momentum} \\
& \frac{\partial P}{\partial t} + \mathbf{u} \cdot \nabla{P} + \rho c^2 \nabla \cdot \mathbf{u} = \frac{1}{\rho} \frac{\beta_v}{c_v \beta_T} (\boldsymbol{\sigma} : \nabla \otimes \mathbf{u} - \nabla \cdot \boldsymbol{q} ) + \alpha_4 + \alpha_5,
\label{eq:pressure}
\end{align}
where $\rho$ is the density, $\mathbf{u}$ is the velocity vector, $P$ is the pressure, $\boldsymbol{\sigma} = \mu \left( \nabla \mathbf{u} + \nabla \mathbf{u}^{T}  \right) - (2 \mu/3)(\nabla \cdot \mathbf{u})\boldsymbol{I}$ is the viscous stress tensor with $\mu$ the dynamic viscosity and $\boldsymbol{I}$ the identity matrix, $c = 1/\sqrt{\rho \beta_s}$ is the speed of sound with $\beta_s=-(1/v)\left(\partial v /\partial P\right)_{s}$ the isentropic compressibility and $v = 1/\rho$ the specific volume, $\beta_v = (1/v)\left(\partial v/\partial T\right)_{P}$ is the volume expansivity with $T$ the temperature, $c_v$ is the isochoric specific heat capacity, $\beta_T = -(1/v)\left( \partial v/\partial P\right)_{T}$ is the isothermal compressibility, and $\boldsymbol{q} = -\kappa \nabla T$ is the Fourier heat conduction flux with $\kappa$ the thermal conductivity.
The terms  $\boldsymbol{\alpha}_1$,  $\alpha_4$ and  $\alpha_5$, as detailed in Section~\ref{sec:SGS_models}, represent the SGS contributions~\cite{Bernades2024-A} of momentum stresses, compressibility effects and heat conduction fluxes, respectively.

\subsection{Real-gas thermodynamics}  \label{sec:real_gas}

The thermodynamic space of solutions for the state variables pressure $P$, temperature $T$ and density $\rho$ of a monocomponent substance is described by an equation of state (EOS).
One popular choice for systems at high pressures, which is used in this study, is the Peng-Robinson~\citep{Peng1976-A} equation of state written as
\begin{equation}
 P = \frac{R_u T}{(M/\rho) - b} -  \frac{a}{(M/\rho)^{2} + 2 b(M/\rho) - b^{2}} + \alpha_6,  \label{eq:peng_robinson}
\end{equation}
where $R_u$ is the universal gas constant and $M$ is the molar mass.
The coefficients $a$ and $b$ take into account real-gas effects related to attractive forces and finite packing volume, respectively, and depend on the critical temperature $T_c$, critical pressure $P_c$, and acentric factor $\omega$.
They are defined as
\begin{align}
 & a = 0.457\frac{\left(\mathrm{R_u} T_{c}\right)^{2}}{P_{c}} \left[ 1 + c \left(1 - \sqrt{T/T_{c}} \right) \right]^{2}, \\
 & b = 0.078\frac{\mathrm{R_u} T_{c}}{P_{c}}, \label{eq:peng_robinson_a_b}
\end{align}
where coefficient $c$ is provided by
\begin{align}
  & c= \left\{ \begin{array}{rl}
                     & 0.380 + 1.485\omega - 0.164\omega^{2} + 0.017\omega^{3} \quad \mbox{ }\mbox{if}\mbox{ }\mbox{ } \omega> 0.49,\\
                     & 0.375 + 1.542\omega - 0.270\omega^{2} \quad\quad\quad\quad\quad \mbox{ }\mbox{ }\mbox{ }\mbox{otherwise}.  \label{eq:peng_robinson_c}
                    \end{array} \right.
\end{align}

The Peng-Robinson real-gas equation of state needs to be supplemented with the corresponding high-pressure thermodynamic variables based on departure functions~\citep{Reynolds2019-B} calculated as a difference between two states.
In particular, their usefulness is to transform thermodynamic variables from ideal-gas conditions (low pressure only temperature dependent) to supercritical conditions (high pressure).
The ideal-gas parts are calculated by means of the NASA 7-coefficient polynomial~\citep{Burcat2005-TR}, while the analytical departure expressions to high pressures are derived from the Peng-Robinson equation of state as detailed in Jofre and Urzay~\cite{Jofre2021-A}.
The low-pass filtering of the equation of state generates an additional unresolved term identified as $\alpha_6$, which is detailed in Section~\ref{sec:SGS_models}.

\subsection{Subgrid scale models}  \label{sec:SGS_models}

\subsubsection{Stress tensor} The SGS stress tensor $\boldsymbol{\alpha}_1 = \rho \boldsymbol{\tau}$ is the most important unclosed term in the momentum equations, which emerges from the non-linearity of the convective term.
This tensor will be assessed by means of the eddy-viscosity concept, where the anisotropic part of the tensor ${\tau_{ij}^a}^{SGS}$ is modeled as
\begin{equation}
{\tau_{ij}^a} = \tau_{ij} - \frac{1}{3} \delta_{ij} \thinspace\tau_{kk} = - \nu_{SGS} \thinspace {S_{ij}(\mathbf{u})},  \label{eq:eddy_viscosity_model}
\end{equation}
where $\nu_{SGS}$ is the turbulent viscosity, $\delta_{ij}$ is the Kronecker symbol, and ${S_{ij}(\mathbf{u})} = \thinspace (\partial \mathbf{u}_i / \partial x_j + \partial \mathbf{u}_j / \partial x_i) - 2/3 \thinspace \delta_{ij} \thinspace \partial \mathbf{u}_k / \partial x_k$ is the rate-of-strain tensor.
It is important to note that, for convention, the Einstein notation is utilized.
In eddy-viscosity-type models, $\nu_{SFS}$ governs the magnitude of the tensor, while instead, its degree of anisotropy and orientation are determined by ${S_{ij}(\mathbf{u})}$, hereinafter expressed also as ${S}_{ij}$.
To this extent, three different models for the SGS stress tensor ${\tau_{ij}^a}$ are considered within this study: Classical Smagorinsky (CS), Dynamic Smagorinsky (DS) and wall-adapting local eddy-viscosity (WALE).
The turbulent viscosity of these three models is expressed as (complete details can be found in~\cite{Bernades2024-A})
\begin{numcases}{}
   \nu_{CS} = ({C_S} \thinspace {{\Delta}})^2 \vert {S({u})} \vert,  \label{eq:Smagorinsky} \\
   \nu_{DS} = ({C_D} \thinspace {{\Delta}})^2 \vert {S({u})} \vert,  \label{eq:Smagorinsky_dynamic} \\
   \nu_{WALE} = ({C_w} \thinspace \Delta)^2 \frac{(1/2 \thinspace{S^d}_{ij} \thinspace 1/2 \thinspace{S^d}_{ij})^{3/2}}{(1/2 \thinspace{S}_{ij} \thinspace 1/2 \thinspace {S}_{ij})^{5/2} + (1/2 \thinspace {S^d}_{ij} \thinspace 1/2 \thinspace {S^d}_{ij})^{5/4}}, \label{eq:WALE}
\end{numcases}
where $\vert {S({u})} \vert = (1/2 \thinspace {S}_{ij} {S}_{ij})^{1/2}$, hereafter written also as $\vert {S} \vert$, $C_S = 0.10$ is the Smagorinsky constant, and $\Delta$ is the mesh size. 
The Dynamic Smagorinsky constant depends on the velocity field in the wall-normal direction.
Moreover, ${S^d}_{ij} =\thinspace ({{A}_{ij}}^2 +  {{A}_{ji}}^2) \thinspace - 2/3 \thinspace \delta_{ij} \thinspace {{A}_{kk}}^2$ is the trace-less symmetric part of the square of the Favre-filtered velocity gradient tensor. 
The constant $C_w$ is derived by assuming that the new model gives the same ensemble-average subfilter kinetic energy dissipation as the classical Smagorinsky, and results for this work to $C_w = 0.15$~\cite{Bernades2024-A}.
Finally, the trace of the tensor $\tau_{kk}$ is constructed according to the model by~\citet{Vreman1994-A} as it provides the higher probability to collapse to filtered DNS based on an \textit{a priori} analysis~\cite{Bernades2024-A}.
Its expression reads
\begin{equation}
    \tau_{kk} = C_w {\Delta}^2 \sum_{i,j} {S}_{ij}^2,
\end{equation}
where $C_w = 0.03$ based on the filtered DNS of the problem of interest~\cite{Bernades2024-A}.

\subsubsection{Subgrid scale model for the pressure transport equation}

\paragraph{Compressibility SGS}

The term $\alpha_4$ is associated with flow dilatation due to compressibility effects and its closing is based on the scale-similarity assumption as a function of ${\rho c^2 \nabla \cdot \mathbf{u}}$ yielding
\begin{equation}
    {\alpha_4} = \frac{1}{K_{\alpha_4}} \left( \overline{{\rho}{c}^2 \nabla \cdot {\mathbf{u}}} - {\overline{\rho}} \thinspace {\breve{c}}^2 \thinspace \nabla \cdot \overline{{\mathbf{u}}} \right),
\end{equation}
where in this work $K_{\alpha_4} = -10$ minimizes the sum of squared errors~\cite{Bernades2024-A}; the overbar $\overline{()}$ refers to the low-pass filtered quantity.

\paragraph{Heat conduction SGS}

The term $\alpha_5$ contains the SGS contribution of the viscous  stress tensor and Fourier heat conduction flux $\boldsymbol{{q}} = f({\rho},{T})$, where ${\beta_v} = f({\rho},{P})$, ${\beta_T} = f({\rho},{P})$ and ${c_v} = f({\rho},{P},{T})$.
Similar to the discussion above, the closure model for this SGS term is based on the scale-similarity assumption and reads
\begin{equation}
   {\alpha_5} = \frac{1}{K_{\alpha_5}} \left[{\overline{\frac{1}{\rho} \frac{\beta_v}{c_v \beta_T} (\boldsymbol{\sigma} : \nabla \otimes \mathbf{u} - \nabla \cdot \boldsymbol{q} )}} - \frac{1}{{\overline{\rho}}} \frac{{\overline{\beta_v}}}{{\overline{c_v}} \overline{{\beta_T}}} (\boldsymbol{\overline{{\sigma}}} : \nabla \otimes \overline{{\mathbf{u}}} - \nabla \cdot \overline{{\boldsymbol{q}}})\right],   
\end{equation}
with $K_{\alpha_5} = 4.0$ adjusted to the problem of interest for this \textit{a posteriori} analysis. 

\subsubsection{Subgrid scale model for the equation of state}

The SGS term $\alpha_6$ results from filtering the nonlinear Peng-Robinson equation of state.
In the framework proposed in this work, pressure is transported instead of total energy to attain pressure-equilibrium-preservation, and therefore the unresolved term of the equation of state is contained in the temperature field in the form $T = f(\rho, P) + \alpha_6$. Note that distinction is made between $T$ and $T(\Psi)$. The former represents the resolved LES field, whereas the latter corresponds to the calculated temperature from resolved LES quantities, in particular $\Psi = (\rho, P)$.
The SGS model for $\alpha_6$ is based on the improved LES assumption (ILA)~\cite{Borghesi2015-A} and is expressed as
\begin{equation}
     T = T ({\Psi}) +  \overbrace{C_p \left( \overline{T} - T ({\overline{\Psi}})\right)}^{\alpha_6}.
\end{equation}
where $C_p$ is the ILA constant coefficient computed based on the least-square method of Germano's identity tensors whose details, formulated on filtered-DNS framework, can be found in~\citet{Borghesi2015-A} and~\citet{Bernades2024-A}.

\subsection{High-pressure transport coefficients}  \label{sec:transport_coefficients}

The high pressures involved in the analyses conducted in this work prevent the use of simple relations for the calculation of the dynamic viscosity $\mu$ and thermal conductivity $\kappa$.
In this regard, standard methods for computing these coefficients for Newtonian fluids are based on the correlation expressions proposed by Chung et al.~\cite{Chung1984-A,Chung1988-A}.
These correlation expressions are mainly function of critical temperature $T_c$ and density $\rho_c$, molecular weight $W$, acentric factor $\omega$, association factor $\kappa_a$ and dipole moment $\mathcal{M}$, and the NASA 7-coefficient polynomial~\citep{Burcat2005-TR}; further details can be found in dedicated works, like for example Poling \textit{et al.}~\cite{Poling2001-B} and Jofre and Urzay~\cite{Jofre2021-A}.
In particular, correlation results of the Chung et al.~\cite{Chung1984-A,Chung1988-A} model with respect to the NIST reference database~\cite{NIST-M} has been carefully validated by~\citet{Bernades2022-A}.

\section{Computational approach} \label{sec:computational_approaches}

This section covers (i) a brief summary of the numerical method utilized, as well as the (ii) wall-resolved and (iii) wall-modeled simulation strategies considered.
The validation of the WRLES and WMLES frameworks for a reference low-pressure isothermal turbulent channel flow case~\cite{Moser1999-A} is provided in Appendix~\ref{sec:Appendix_A}.

\subsection{Numerical method}  \label{sec:numerical_scheme}

Simulations of high-pressure transcritical turbulence are strongly susceptible to numerical instabilities due to the presence of nonlinear thermodynamic phenomena and large density gradients.
These instabilities can trigger spurious pressure oscillations that may contaminate the solution and even lead to its divergence.
Consequently, it is highly beneficial that the numerical schemes utilized, in addition to being kinetic-energy-preserving (KEP), also attain the so-called pressure-equilibrium-preserving (PEP) property~\cite{Shima2021-A, Lacaze2019-A}, which consists in being able to maintain a constant pressure and velocity field when these are initially constant.
The numerical scheme utilized in this work has been developed specifically to be simultaneously KEP and PEP. For compressible flow, a family of KEP formulations for the convective term has been recently derived~\cite{Coppola2019-A}. Instead, the latter property is achieved by solving a pressure evolution equation. A thorough description and validation of this method can be found in Bernades \textit{et al.}~\cite{Bernades2022b-A,Bernades2022d-A,Bernades2023b-A}.
In this regard, recent \textit{a priori} analysis~\cite{Bernades2024-A} assessed this numerical method to characterize the SGS terms, and subsequently proposes a closure framework suitable to large-eddy simulations of wall-bounded high-pressure transcritical fluid turbulence.

\subsection{Wall-resolved simulations}  \label{sec:wall_resolved}

Wall-resolved LES is conducted with the problem of interest described in Section~\ref{sec:DNS_setup}. The flow is controlled by means of a proportional controller imposed as a volumetric force to guarantee the same bulk velocity as the reference DNS.
The grid resolution selected corresponds to $48 \times 48 \times 48$ points and is stretched towards the walls in the vertical direction with the first grid point at $y^{+} = y u_{\tau,w}/\nu_{w} = 1$ with respect to each wall. The resulting resolutions range $1.0 \le \Delta y_{cw}^{+} \le 3.3$ and $1.0 \le \Delta y_{hw}^{+} \le 7.8$ for the cold and hot walls, respectively. Instead, it is uniform in the streamwise and spanwise directions with resolutions in wall units (based on $cw$ values) equal to $\Delta x^{+} = 15.0$ and $\Delta z^{+} = 5.0$.

The WRLES approach relies solely on SGS models (see Section~\ref{sec:SGS_models}) to represent the unresolved small flow scales. In particular, the CS, DS and WALE models are assessed for $\boldsymbol{\alpha_1}$ along the closure expressions for $\alpha_4$ and $\alpha_5$, whose influence were found to be important.
Nevertheless, it is known from previous \textit{a priori} analysis that neither of these SGS stress tensor models were able to recover the small-eddy properties in terms of magnitude, shape and orientation~\cite{Jofre2018-A,Jofre2019-A,Bernades2024-A} of the filtered DNS.
Additionally, it is noted that the ILA model proposed for closing the nonlinear equation of state has resulted in misleading outputs. In fact, it contaminates the thermodynamic fields excessively, and therefore it is excluded from the present \textit{a posteriori} analysis.
To this aim, the WRLES computation results are summarized in Section~\ref{sec:results_wall_resolved} and the heat-transfer analysis in Section~\ref{sec:heat_transfer}. Further details of the effects of the SGS models for the pressure transport equation are briefly covered in Appendix~\ref{sec:Appendix_B}.

\subsection{Wall-modeled simulations}  \label{sec:wall_modeled}

Fully resolving the boundary layers in LES approaches is typically not affordable. Consequently, the wall laws are leveraged to model wall fluxes, and therefore reduce the expensive computational costs near the wall which limits the efficiency of DNS and WRLES approaches.
In particular, the wall region is not resolved and the wall fluxes cannot be directly evaluated. Instead, the wall fluxes (wall shear stress and Fourier flux) algebraically link the velocity at some distance from the wall, and to this end, such strategy reduces the computational cost of resolving the flow motion in the vicinity of walls. However, the use of simple laws of the wall are not fully satisfactory~\cite{Potier2018-B}, and specific models are necessary to accurately recover the modeled part of the domain.

\begin{figure}
	\centering
     {\includegraphics[width=0.85\linewidth]{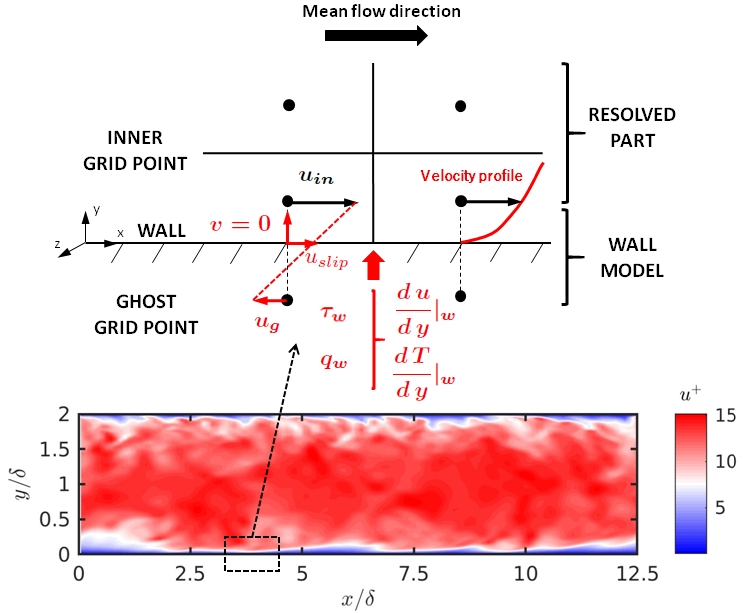}} 
 \vspace{1mm}
	\caption{Wall-modeled discretization approach and near-wall velocities and temperatures scheme for the problem of interest depicted in Figure~\ref{fig:DNS_setup}.} \label{fig:WMLES_mesh_setup}
\end{figure}

In this regard, current wall models may capture and predict a wide range of flow phenomena, such as heat fluxes, streamwise and adverse pressure gradients, non-equilibrium effects and/or large temperature gradients and chemical reactions.
This study is, however, limited to two wall models currently utilized for wall-bounded high-pressure transcritical fluid systems: a (i) dynamic model based on the ``standard law of the wall'' and (ii) a recently developed coupled model~\cite{Cabrit2009-B,Cabrit2009-A}, where the velocity and temperature fields are linked by means of an optimization algorithm.
This latter model is known to correctly predict the solution for isothermal high-pressure transcritical turbulent channel flow.
However, in this work, the model will be stressed by making it operate at relatively low-Reynolds-number conditions such that the thermodynamic effects become comparable to the shear stresses from the walls.
Hence, the objectives of the WMLES computations are twofold: (i) further reduce the mesh size to accelerate the computations compared to WRLES, and (ii) provide robustness to the heat-transfer related quantities based on the direct modeling of the wall fluxes for the shear stress and the Fourier heat transfer.
In this case, the grid is stretched toward the walls in the vertical direction with the first grid point at $y^{+} = y u_{\tau,w}/\nu_{w} = 30$ (log-law region)~\cite{Cabrit2009-B}.
In addition, the grid is uniform within the inner region with similar resolutions as the WRLES case and corresponding to $\Delta y_{cw}^{+} = 2.5$ and $\Delta y_{hw}^{+} = 12.6$. The mesh in the streamwise and spanwise directions is uniform with resolutions in wall units (based on $cw$ values) equal to $\Delta x^{+} = 13.2$ and $\Delta z^{+} = 4.4$, respectively.
Thus, this arrangement corresponds to a grid size of $48 \times 30 \times 48$ points.
Finally, Figure~\ref{fig:WMLES_mesh_setup} depicts the discretization approach and the near-wall velocity and temperature methodology for WMLES strategies based on the compressible solver and numerical method employed.

\subsubsection{Dynamic wall model}

This model is based on the ``standard law of the wall''~\citep{VanDriest1951-A} to predict wall shear stresses and heat fluxes, and consequently recover the velocity and temperature of the inner part of the domain.
To predict the wall fluxes, the model needs to first solve for the friction velocity and temperatures, denoted by $()_\tau$, yielding
\begin{align}
    u_\tau = \sqrt{\frac{\tau_w}{\rho_w}}, \quad \quad T_\tau = \frac{q_w}{\rho_w {c_p}_w u_\tau}, \label{eq:normalization_tau}
\end{align}
where sub-index $()_w$ corresponds to the wall quantity. Based on these expressions, the wall-normal distance, velocity and temperature can be expressed in wall units (super-index notation $()^+$) as
\begin{align}
    y^+ = \frac{y \rho_w u_\tau}{\mu_w}, \quad \quad u^+ = \frac{u}{u_\tau}, \quad \quad T^+ = \frac{T_w - T}{T_\tau}. \label{eq:normalization_wall}
\end{align}
Such normalization allows to express the ``standard law of the wall'' for the velocity as
\begin{numcases}{}
   u^+ = y^+,  \quad \quad \quad \quad \quad \mbox{in viscous region} \mbox{}, y^+ < y^+_t, \\
   u^+ = \frac{1}{\kappa_u} \textrm{ln} \thinspace y^+ + C_u, \mbox{ in inertial log-law region} \mbox{}, y^+ \ge y^+_t, \label{eq:dynamic_velocity_law} 
\end{numcases}
where $\kappa_u = 0.41$ and $C_u = 5.5$ are the von Karman coefficient and Van Driest constant, respectively, for internal flows, and $y^+_t = 11.5$ defines the intersection between the viscous sublayer and log-law region.
Analogously, the thermodynamic (temperature) law can be expressed as
\begin{numcases}{}
   T^+ = Pr y^+,   \quad \quad \quad \quad \mbox{ } \mbox{in viscous region} \mbox{}, y^+ < y^+_t, \label{eq:dynamic_temperature_law} \\
   T^+ = \frac{{Pr}_t}{\kappa_u} \textrm{ln} \thinspace y^+ + \beta,\mbox{ in inertial log-law region} \mbox{}, y^+ \ge y^+_t, \label{eq:dynamic_temp_law} 
\end{numcases}
where the turbulent Prandtl number $Pr_t = 0.85$ and the integration constant $\beta$ depend on the Prandtl number as proposed by~\citet{Kader1981-A} in the form
\begin{equation}
    \beta = \left( 3.85 {Pr}^{1/3} - 1.3 \right)^2 + 2.12 \thinspace \textrm{ln} (Pr).
\end{equation}

The implementation in the solver utilizes, however, empiric velocity~\cite{Reichardt1956-A} and temperature~\cite{Kader1981-A} laws to avoid numerical instabilities and bending between the viscous and log-law regions.
These expressions read as
\begin{align}
    u^+ & = \frac{1}{\kappa_u} \textrm{ln} (1 + \kappa_u y^+) + 7.8 \left( 1 - e^{-y^+/11} - \frac{y^+}{11}e^{-y^+/3}\right), \label{eq:u_plus_empirique} \\
    T^+ & = Pr \thinspace y^+ e^{\Gamma} + \left[ 2.12 \textrm{ln} (1 + y^+) + C_T \right] e^{1/\Gamma}, \label{eq:T_plus_empirique}
\end{align}
with the integration constant $C_T = (3.85 Pr^{1/3} - 1.3)^2 + 2.12 \textrm{ln} Pr$ and function $\Gamma = - 10^{-2} (Pr y^+)^4 / (1 + 5 Pr^3 y^+)$.
Therefore, by inverting these equations and solving for $u_\tau$ and $T_\tau$, the wall fluxes can be then obtained along with the wall quantities as
\begin{numcases}{}
     \tau_w = \rho_w \thinspace {u_\tau}^2, \label{eq:shear_flux} \\
     q_w    = T_\tau \thinspace \rho_w \thinspace {c_p}_w \thinspace u_\tau. \label{eq:heat_flux}
\end{numcases}

\subsubsection{Coupled wall model}

The overarching objective of this coupled model is to consider the strong thermodynamic (temperature) gradients between the wall and the first inner point resolved by the computations~\cite{Cabrit2009-B}.
This model is expressed as
\begin{numcases}{}
\begin{cases}{}
     u^+ = y^+,  \quad \quad \quad \quad \quad \quad \quad \quad \quad \quad \quad \quad \quad \quad \quad \quad  \quad y^+ < y^+_t, \\
       \frac{2}{Pr_t B_q} \left( \sqrt{1 - K \thinspace B_q} - \sqrt{\frac{T}{T_w}}  \right) = \frac{1}{\kappa_u} \textrm{ln} \thinspace y^+ + C_u, \quad y^+ \ge y^+_t, \label{eq:coupled_velocity_law}  \\

\end{cases} \\
       T^+ = {Pr} \thinspace y^+ e^{\Gamma} + \left( {Pr}_t \thinspace u^+ + K \right) e^{1/\Gamma}, \label{eq:coupled_temp_law} 
\end{numcases}
where $B_q = T_\tau / T_w$ and $K$ and $\Gamma$ are constants depending on the Prandtl number as
\begin{numcases}{}
    K = \beta - {Pr}_t \thinspace C_u + \left( \frac{{Pr}_t}{\kappa_u} - 2.12 \right) \left( 1 - 2 \thinspace \textrm{ln} \thinspace (20)\right), \\
    \Gamma = - \frac{10^{-2} (Pr \thinspace y^+)^4}{1 + 5 \thinspace {Pr}^3 \thinspace y^+}.
\end{numcases}

In the viscous sub-region, Eq.~\ref{eq:coupled_velocity_law} collapses to Eq.~\ref{eq:dynamic_velocity_law}.
Therefore, the inversion of this system can be solved for $u_\tau$ and $T_\tau$ by properly optimizing the non-liner system.
Knowing these quantities based on the wall and first inner point, the wall fluxes can be analogously derived from the ``standard law of the wall''. 
Consequently, the system of Eqs.~\ref{eq:coupled_velocity_law}-\ref{eq:coupled_temp_law} is rewritten as
\begin{numcases}{}
\begin{cases}{}
     u_\tau = \sqrt{\frac{u_{in} \mu_w}{y_{in} \rho_w}},  \quad \quad \quad \quad \quad \quad \quad \quad \quad \quad \quad \quad \quad \quad \quad \quad \quad \quad \quad \quad \quad \quad \quad \quad \mbox{\space} \\
       \frac{2 T_w (Pr_t u_{in} + K u_\tau)}{Pr_t (T_w - T_{in})} \left( \sqrt{1 - \frac{K (T_w - T_{in})}{Pr_t u_{in} + K u_\tau} \thinspace u_\tau} - \sqrt{\frac{T}{T_w}}  \right) = \left[ \frac{1}{\kappa_u} \textrm{ln} \thinspace (\frac{y_{in} \rho_w}{\mu_w}) + C_u \right] u_\tau, \label{eq:coupled_velocity_law_inv}  \\

\end{cases} \\
       \frac{T_w - T_{in}}{T_\tau} = Pr \frac{\rho_w y_{in} u_\tau}{\mu_w} \thinspace e^{\Gamma} + \left( {Pr}_t \thinspace \frac{u_{in}}{u_\tau} + K \right) e^{1/\Gamma}, \label{eq:coupled_temp_law_inv} 
\end{numcases}
to solve for $u_\tau$ (first or second part of Eq.~\ref{eq:coupled_velocity_law_inv} whether $y^+ < y^+_t$ or $y^+ \ge y^+_t$, respectively), and consequently $T_\tau$ can be obtained from the wall shear stress and heat flux (Eq.~\ref{eq:shear_flux}-\ref{eq:heat_flux}), identically as for the dynamic model, where sub-index $()_{in}$ refers to the first inner grid point.

\section{Large-eddy simulation results} \label{sec:results_LES}

The SGS stress tensor models introduced in Section~\ref{sec:flow_physics_modeling} are assessed for both wall-resolved and wall-modeled approaches. The problem of interest is summarized below and the LES computations are obtained utilizing the in-house flow solver RHEA~\cite{RHEA2023-A,Abdellatif2023-A}.

\subsection{Benchmark DNS case} \label{sec:DNS_setup}

\begin{figure}
	\centering
	{\includegraphics[width=0.99\linewidth]{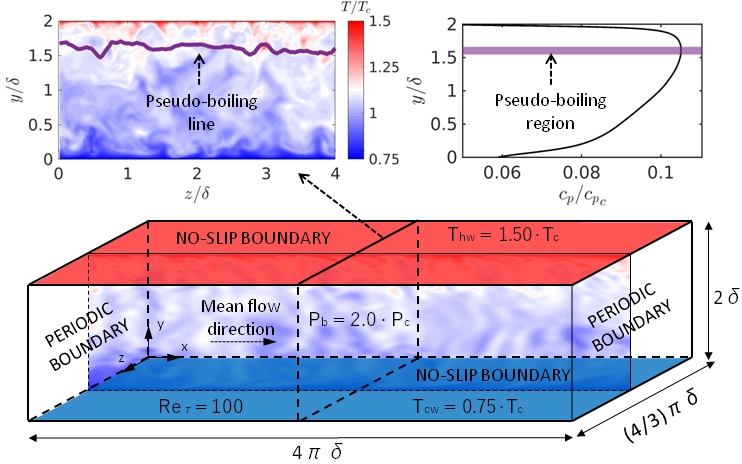}} 
 \vspace{1mm}
	\caption{Schematic of the channel flow problem studied highlighting the operating region of the pseudo-boiling line.} \label{fig:DNS_setup}
\end{figure}

The DNS of transcritical channel flow operates with N$_2$ at a supercritical bulk pressure of $P_b/P_c = 2$ and confined between bottom/cold ($cw$) and top/hot ($hw$) isothermal walls, separated in this case at a distance $H = 2\delta$ with $\delta=100\thinspace{\mu \textrm{m}}$ the channel half-height, at $T_{cw}/T_c = 0.75$ and $T_{hw}/T_c = 1.5$, respectively, where sub-indexes $()_{b}$ and $()_{c}$ correspond to bulk and critical point values. This configuration forces the fluid to undergo a transcritical trajectory by operating within a thermodynamic region across the pseudo-boiling line. In this regard, a schematic of the problem of interest is displayed in Figure~\ref{fig:DNS_setup}.
The  friction Reynolds number at the cold wall is set to $\text{Re}_{\tau,cw} = \rho_{cw} u_{\tau,cw} \delta / \mu_{cw} = 100$ to ensure fully-developed turbulent flow conditions~\cite{Bernades2022-A}. Based on the fully-resolved DNS dataset~\cite{Bernades2023a-A}, the corresponding dimensional parameters are: dynamic viscosity $\mu_{cw} = 1.6\cdot10^{-4}\thinspace\textrm{Pa}\cdot\textrm{s}$, density $\rho_{cw} = 839.4\thinspace\textrm{kg/m}^\textrm{3}$, and friction velocity $u_{\tau,cw} = 1.9\cdot10^{-1}\thinspace\textrm{m/s}$.
The boundary conditions imposed correspond to an impermeable no-slip condition for the wall normal-direction, and periodic for the streamwise and spanwise directions.

The computational domain is $4 \pi \delta \times 2\delta \times 4/3\pi \delta$ in the streamwise ($x$), wall-normal ($y$), and spanwise ($z$) directions, respectively.
The grid is uniform in the streamwise and spanwise directions with resolutions in wall units (based on $cw$ values) equal to $\Delta x^{+} = 9.8$ and $\Delta z^{+} = 3.3$, and stretched toward the walls in the vertical direction with the first grid point at $y^{+} = y u_{\tau,cw}/\nu_{cw} = 0.1$ corresponding to wall resolution ranges $0.1 \le \Delta y_{cw}^{+} \le 2.2$ and $0.1 \le \Delta y_{hw}^{+} \le 3.9$ for the cold and hot walls, respectively.
The simulation strategy starts from a linear velocity profile with random fluctuations, which is advanced in time to reach turbulent steady-state conditions after approximately five flow-through-time (FTT) units; based on the bulk velocity $u_{b}$ and the length of the channel $L_x = 4\pi\delta$, a FTT is defined as $t_{b} = L_x/u_{b} \sim \delta/u_{\tau}$.
In this regard, flow statistics are collected for roughly $10\times$ FTTs once steady-state conditions are achieved. The numerical method used to solve the governing equations is described in Section~\ref{sec:numerical_scheme}.

\subsection{WRLES results} \label{sec:results_wall_resolved}

The WRLES computations analyze the difference between three SGS stress tensors (CS, DS and WALE) assessed against the fully-resolved DNS result.
In particular, this work, as a first LES exploration of the framework presented, focuses on the \textit{a posteriori} assessment of first-order flow statistics; second-order statistics and energy budgets are not characterized in the numerical solution yielding large difference in turbulent fluctuations compared to the reference solution~\cite{Gullbrand2003-A, Mukha2015-TR, Jofre2018-A}. In this regard, modelling the energy corresponding to these fluctuations will be considered in future works.
Moreover, the effects of the SGS term of the pressure transport equation on the first-order flow statistics are detailed in Appendix~\ref{sec:Appendix_B}.
In detail, the $\alpha_4$ and $\alpha_5$ terms are considered in the results presented.
However, the SGS term of the equation of state, denoted as $\alpha_6$, is not included in the present computations as its effect on the flow field is excessively impacting the solution, and consequently its consideration is deferred to future work.

\begin{figure}
	\centering
	\subfloat[]{\includegraphics[width=0.48\linewidth]{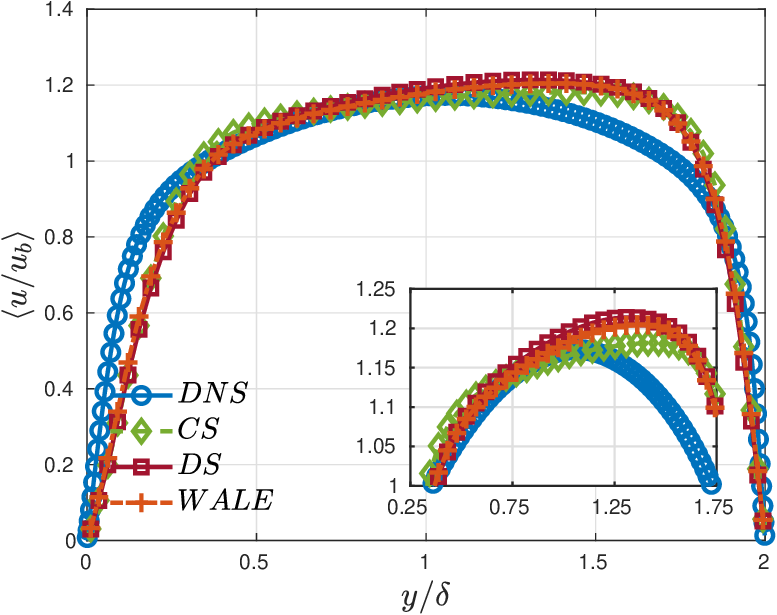}} \hspace{0.5mm}
    \subfloat[]{\includegraphics[width=0.505\linewidth]{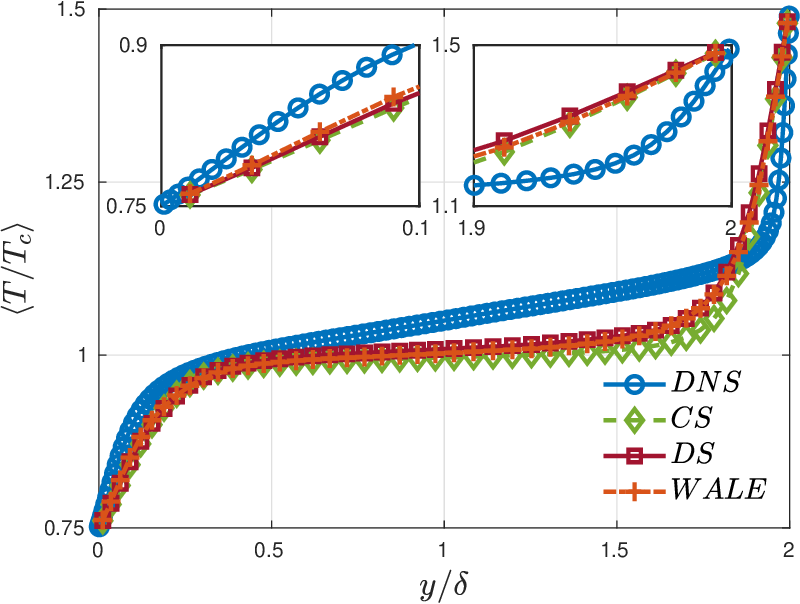}} \\
	\caption{Time-averaged velocity (a) and temperature (b) profiles in the wall-normal direction normalized by channel half-height ($y/\delta$) comparing WRLES results with respect to the fully-resolved DNS dataset.} \label{fig:u_T_vs_y_WRLES}
\end{figure}

\begin{figure}
	\centering
	\subfloat[]{\includegraphics[width=0.48\linewidth]{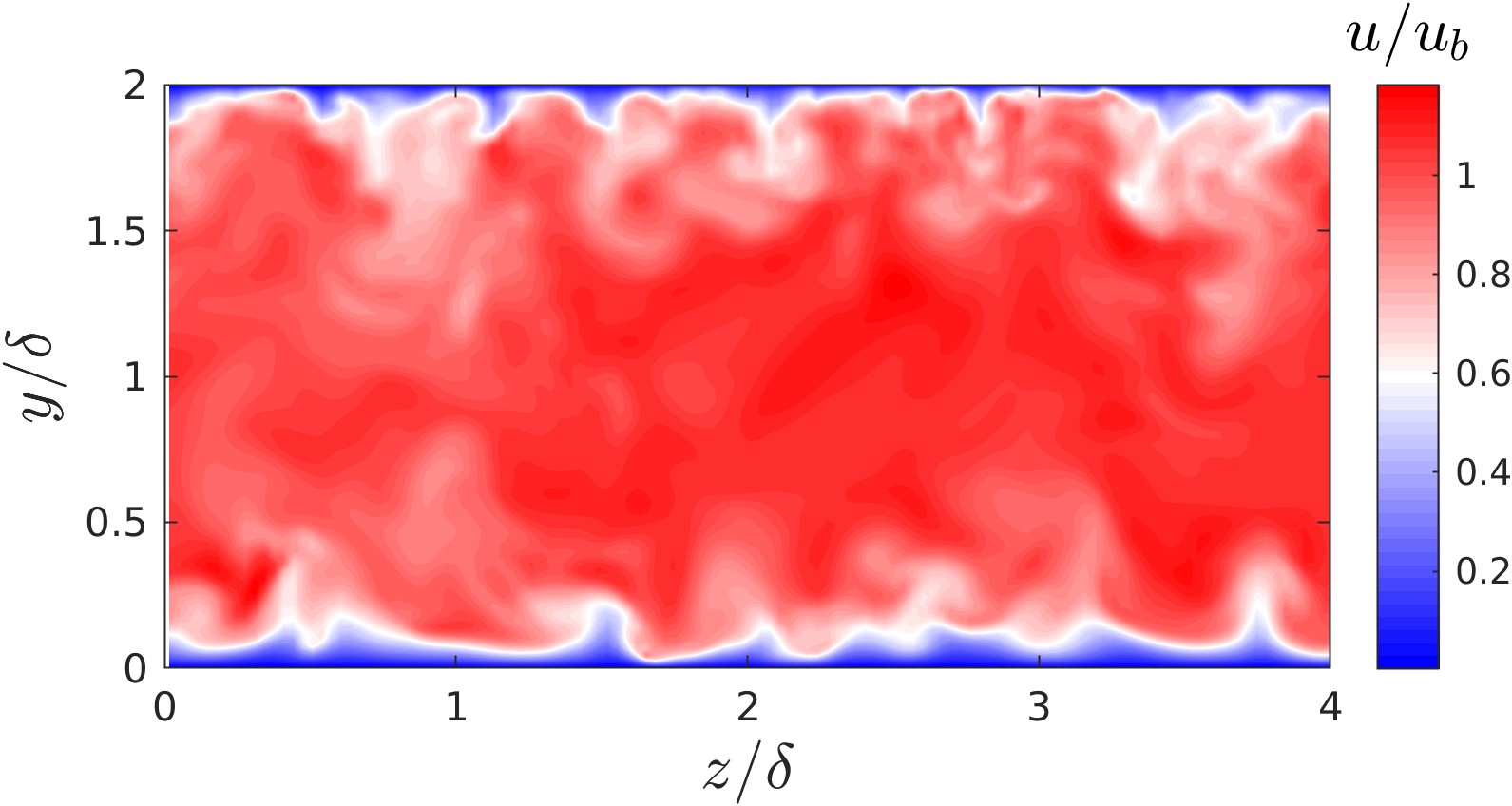}}
    \subfloat[]{\includegraphics[width=0.49\linewidth]{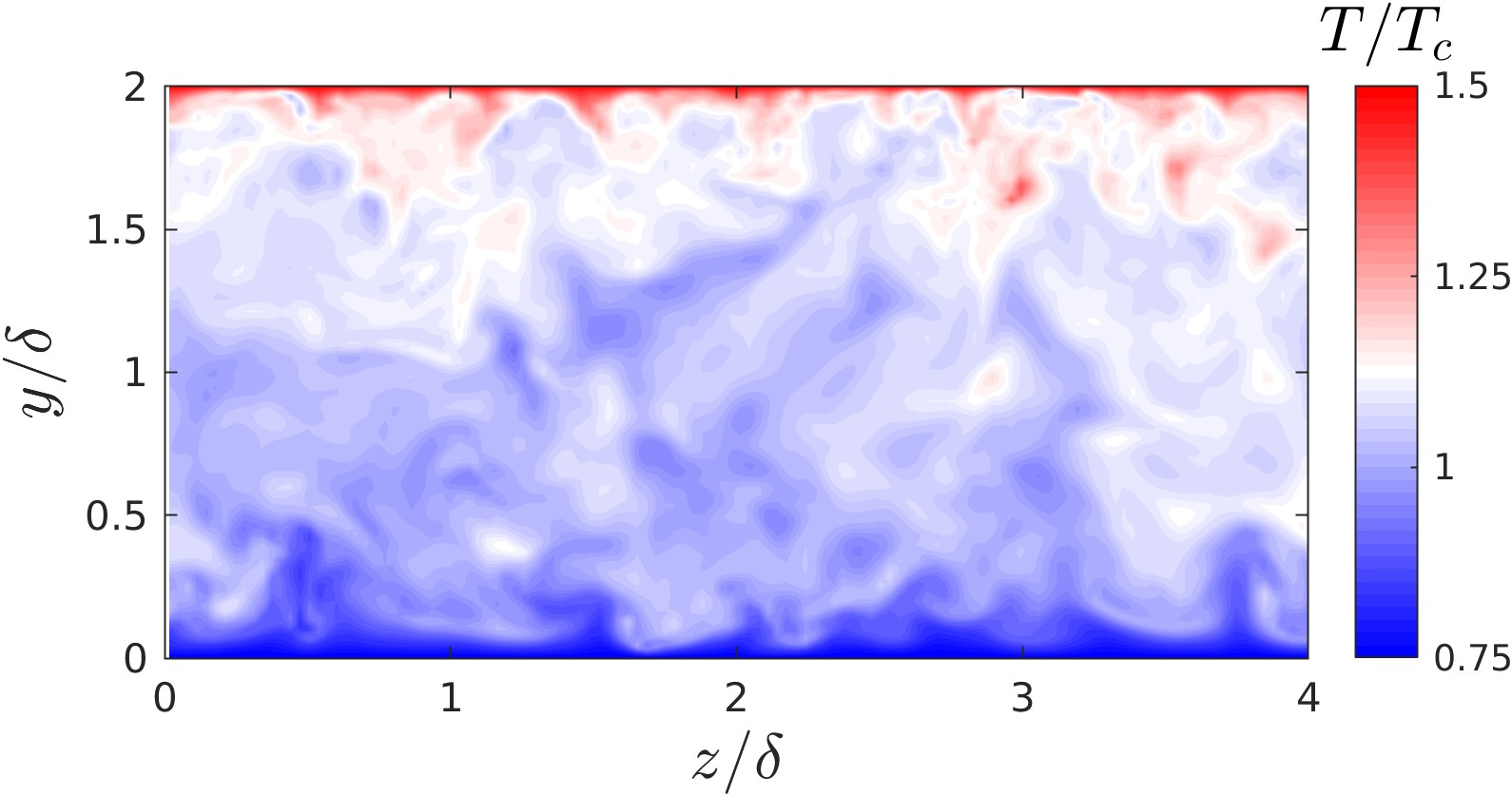}} \\
    \subfloat[]{\includegraphics[width=0.49\linewidth]{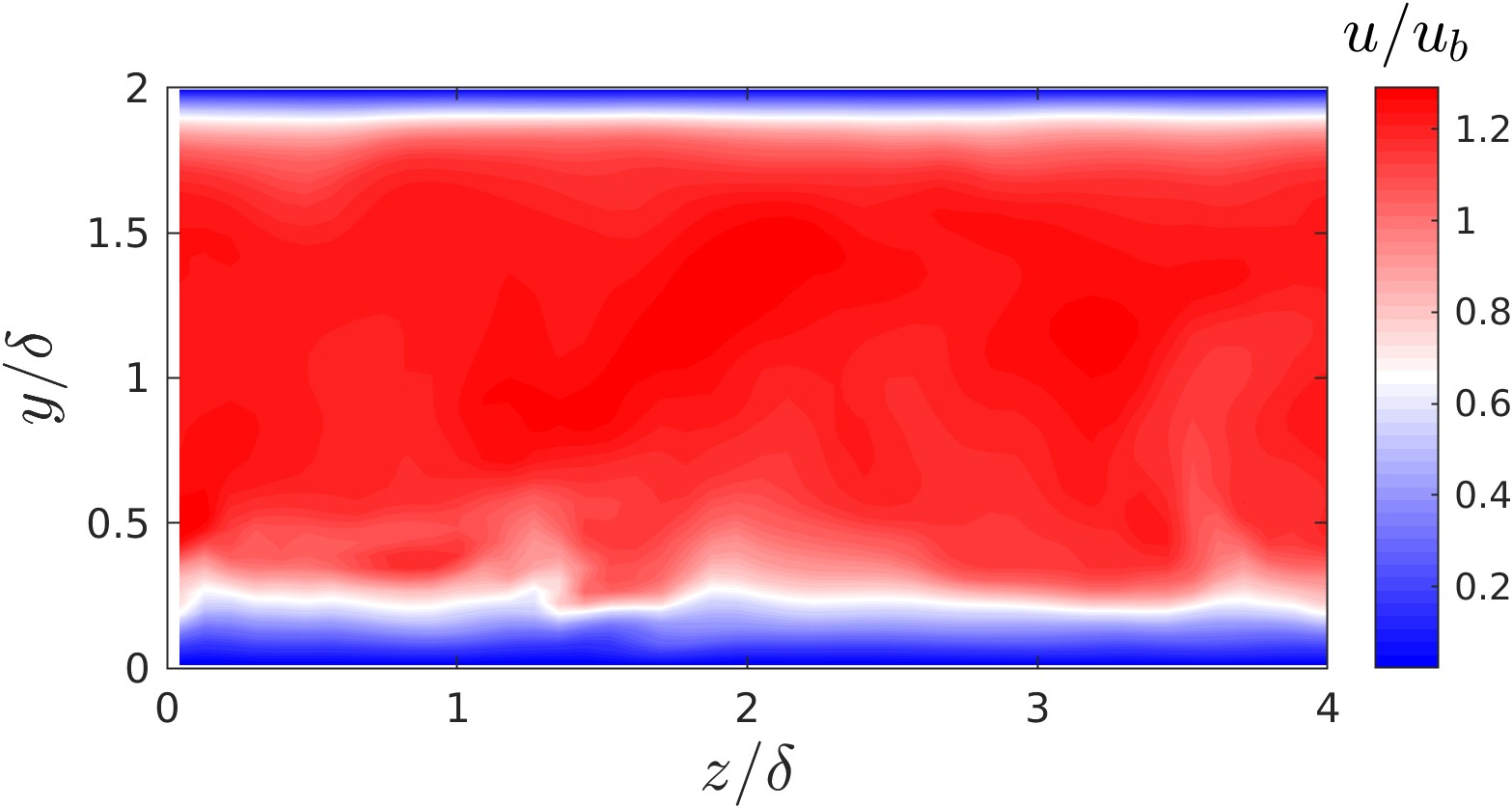}}
    \subfloat[]{\includegraphics[width=0.49\linewidth]{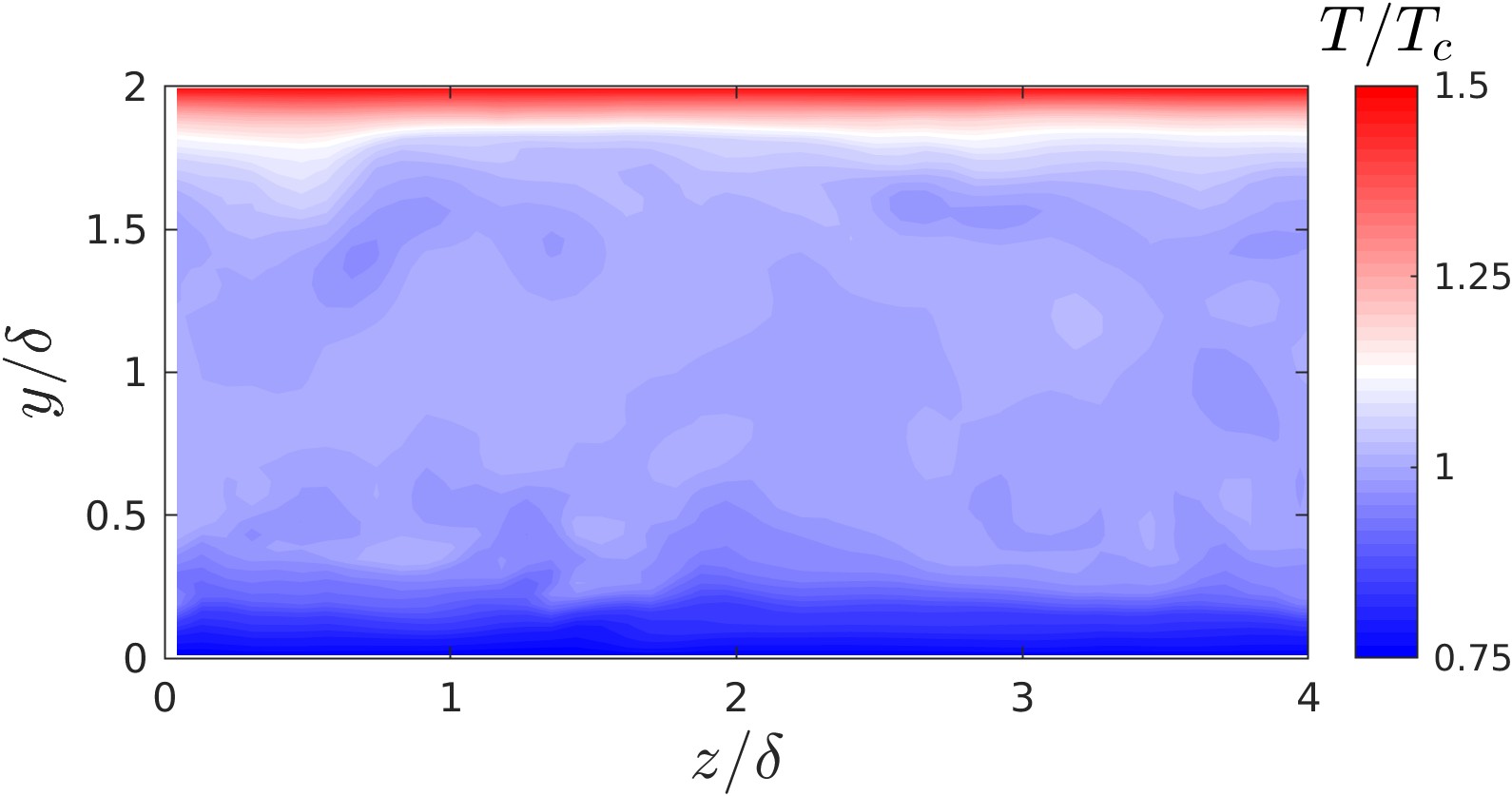}} \\
    \subfloat[]{\includegraphics[width=0.49\linewidth]{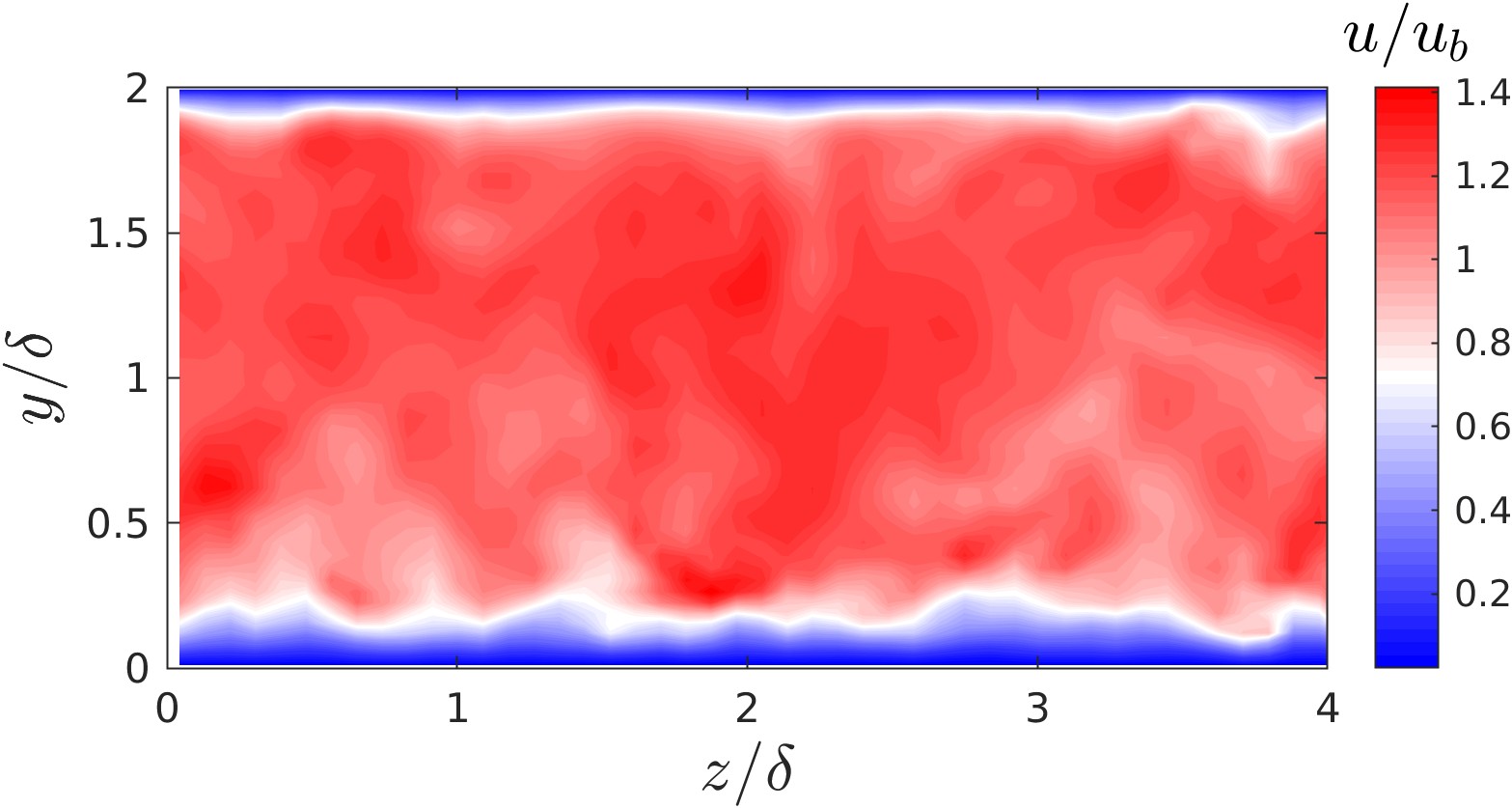}}
    \subfloat[]{\includegraphics[width=0.49\linewidth]{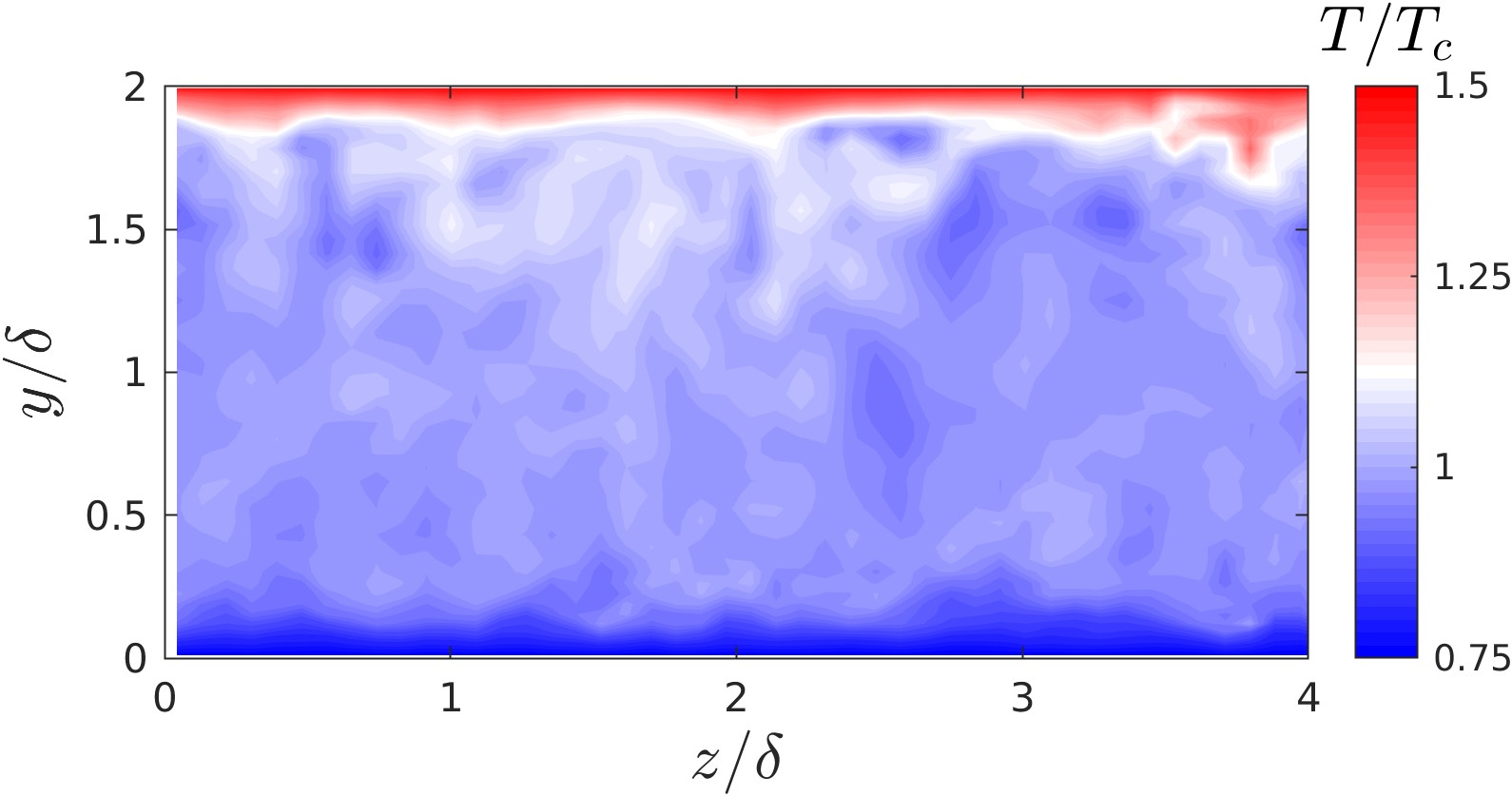}} \\
    \subfloat[]{\includegraphics[width=0.49\linewidth]{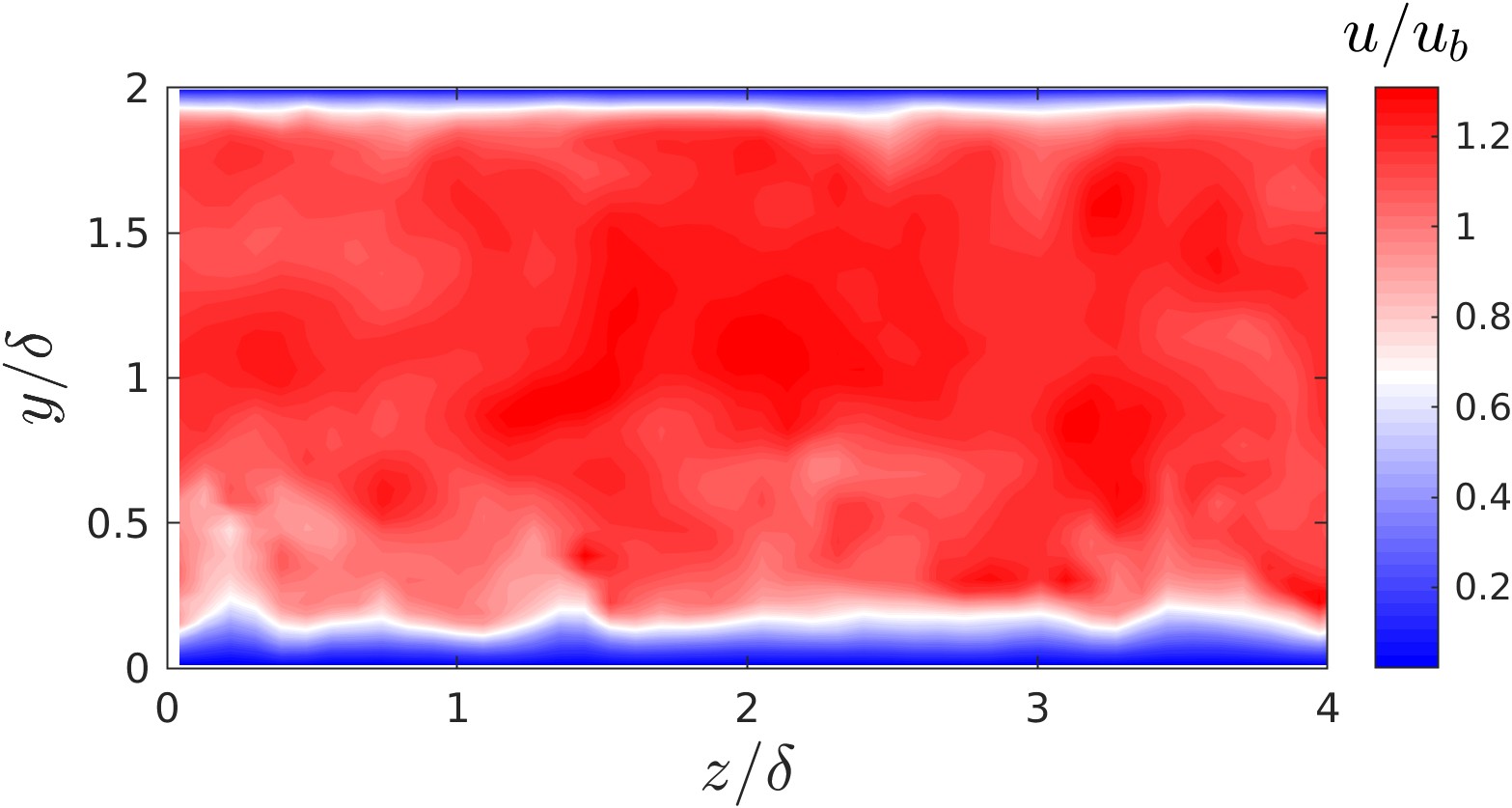}}
    \subfloat[]{\includegraphics[width=0.49\linewidth]{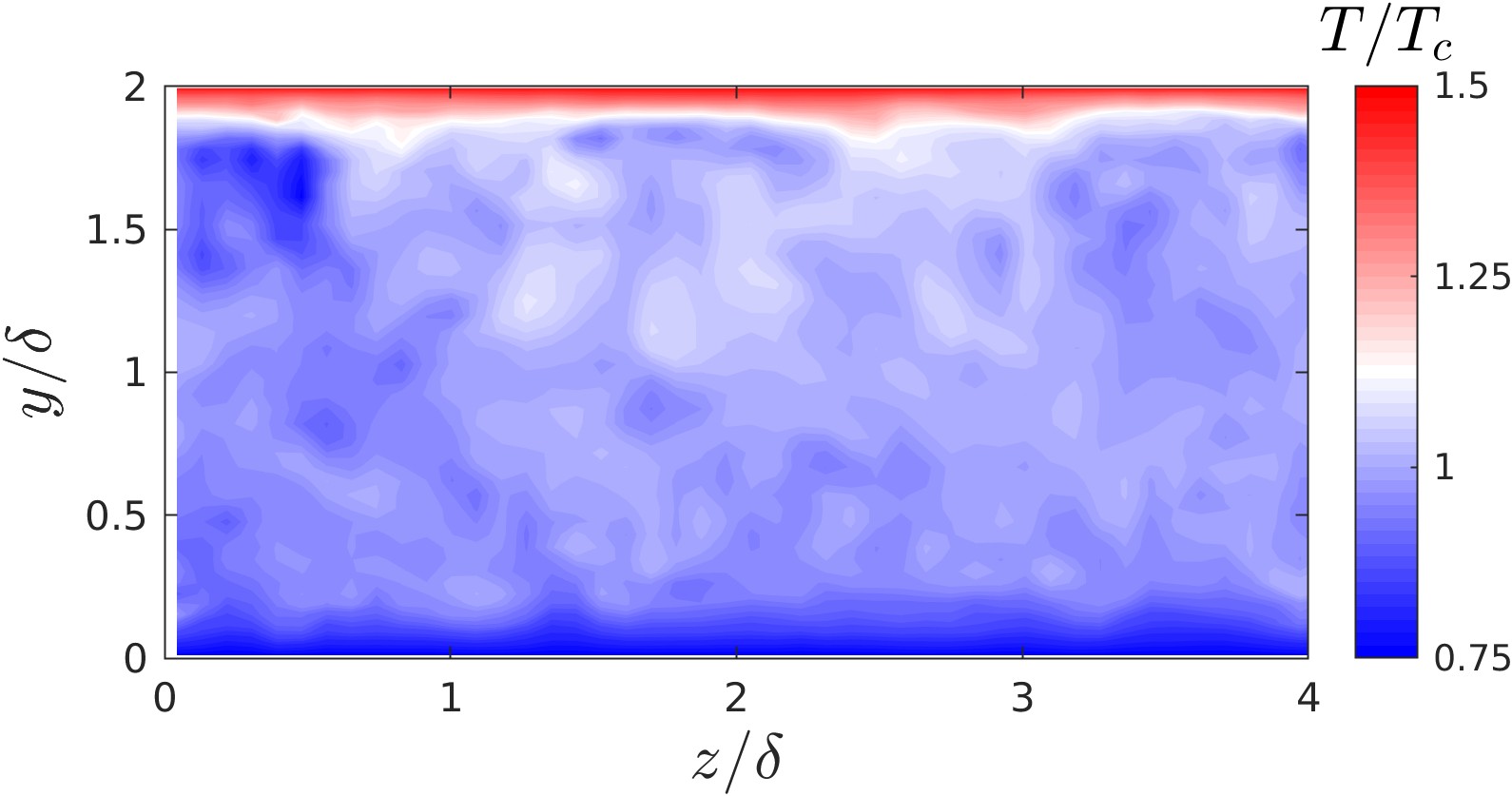}}
	\caption{Snapshots along the spanwise and wall-normal directions of velocity in \review{outer scales} (a,c,e,g) and temperature normalized to the critical point (b,d,f,h) comparing DNS (a,b) the results of the WRLES cases: CS (c,d), DS (e,f) and WALE (g,h).} \label{fig:snapshots_spanwise_DNS_WRLES}
\end{figure}

The mass flow rate in the streamwise direction is imposed through a body force controlled by a feedback loop, which in this case aims at reducing the difference between the target (DNS) and measured bulk velocity.
The first-order flow statistics expressed in outer scales are depicted in Figure~\ref{fig:u_T_vs_y_WRLES}.
In particular, the time-averaged velocity and temperature profiles along the wall-normal direction for the results of the three SGS stress tensors and the fully resolved DNS~\cite{Bernades2023a-A} are presented. 
It is indeed observed that neither SGS model collapses to the reference result.
Only, at the center of the channel the profiles recover similar velocities than the DNS.
Moreover, the CS result is biased towards slightly larger velocities at the center of the channel, which is indicative of laminarization trends, yielding a more parabolic-like profile.
In terms of temperature, the profiles of the WRLES cases achieve similar shapes as the DNS at the cold wall, but towards the center and near the hot wall, the WRLES cases fail to properly represent the DNS result.
It is important to note that Figure~\ref{fig:statistics_WMLES} shows the inner-scaling results for the CS case.
It is seen that at the cold wall the models overpredict the velocity in the log-law region, while an opposite trend occurs at the hot wall.
As expected, the temperature profile obtained deviates for at walls.
In addition, instant snapshots of velocity and temperature are qualitatively compared in Figure~\ref{fig:snapshots_spanwise_DNS_WRLES}.
It is noted that, indeed, the WRLES cases do not properly represent the turbulence intensity and strong temperature gradients of the DNS.
In this regard, an extended analysis of the heat transfer performance is assessed in Section~\ref{sec:heat_transfer}.


\subsection{WMLES results} \label{sec:results_wall_modeled}

This section presents the WMLES results for the (i) dynamic/standard (WMLES-D) and (ii) coupled (WMLES-C) ``standard law of the wall'' models.
For the sake of clarity, only results of the SGS stress tensor model with best performance (CS) is presented.
For completeness, Appendix~\ref{sec:Appendix_C} covers the performance comparison of CS with respect to DS and WALE.
In this case, the mass flow rate is controlled to achieve the same shear-stress level as the DNS.
Previous research~\cite{Bernades2023a-A} has shown that the ``standard law of the wall'' does not agree with the DNS computations when operating under transcritical regimes with strong density gradients.
In particular, the enhanced vorticity in the vicinity of the hot wall, which corresponds to the pseudo-boiling region, due to a baroclinic torque is a fundamental mechanism driving the flow physics in such type of flows.
Consequently, the flow behaviour deviates from standard isothermal wall-bounded turbulence.
In fact, this is a well-documented phenomenon in supercritical fluids, the so-called heat transfer deterioration, which occurs when the fluid operates near the pseudo-boiling line and the wall temperature is above the critical temperature~\cite{Shiralkar1968-TR}.
To this extent, unlike WMLES-D, the WMLES-C derives the velocity value such that non-constant temperature effects are considered.
Therefore, this model typically becomes a well-suited candidate for velocity and temperature prediction under nonlinear thermodynamic regimes. Despite these efforts, this model is still limited to the ``standard law of the wall'' model, which has not yet been extended to wall-bounded high-pressure transcritical turbulent flows.

\begin{figure}
	\centering
	\subfloat[]{\includegraphics[width=0.49\linewidth]{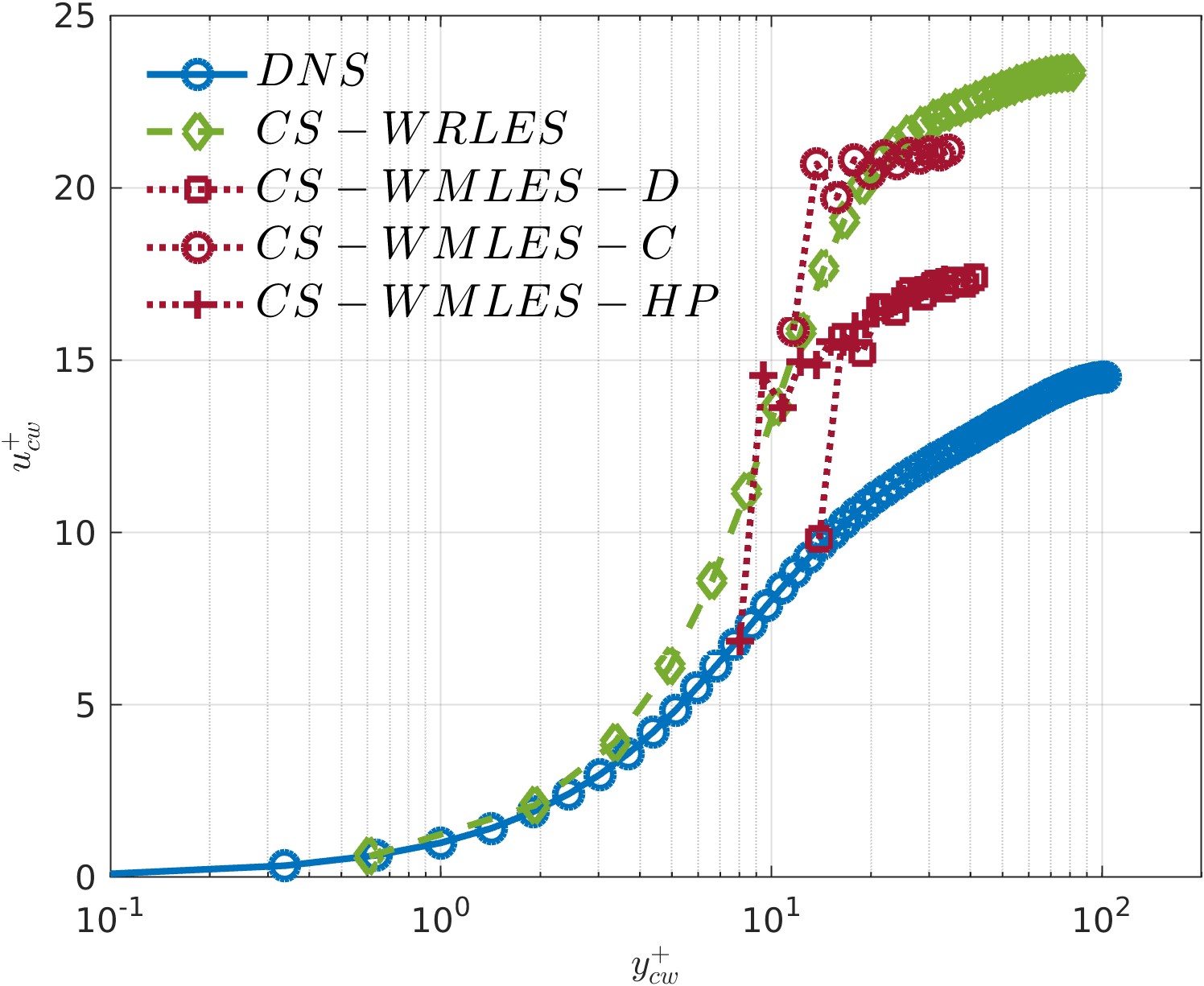}} \hspace{0.5mm}
    \subfloat[]{\includegraphics[width=0.49\linewidth]{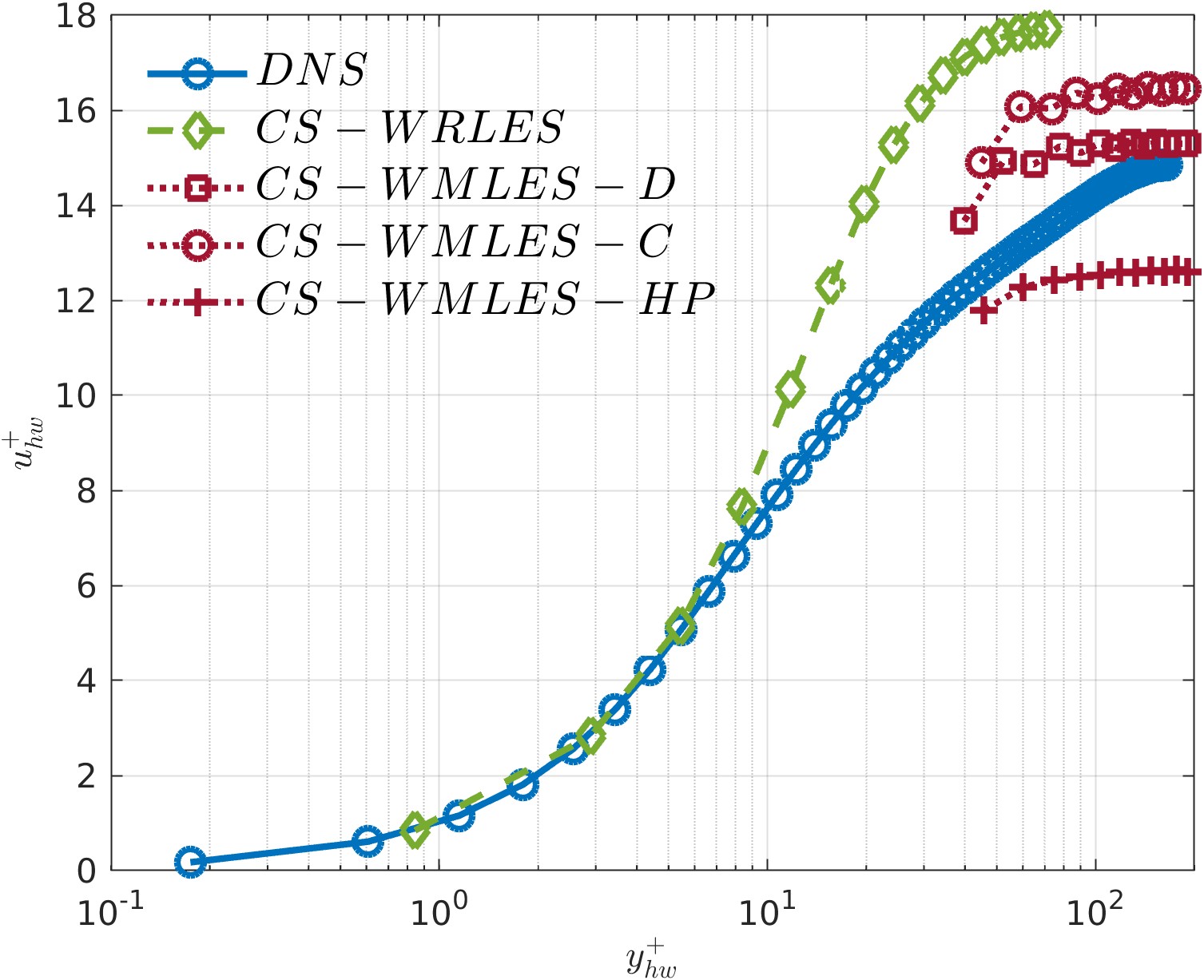}} \\ \vspace{0.9mm}
    \subfloat[]{\includegraphics[width=0.48\linewidth]{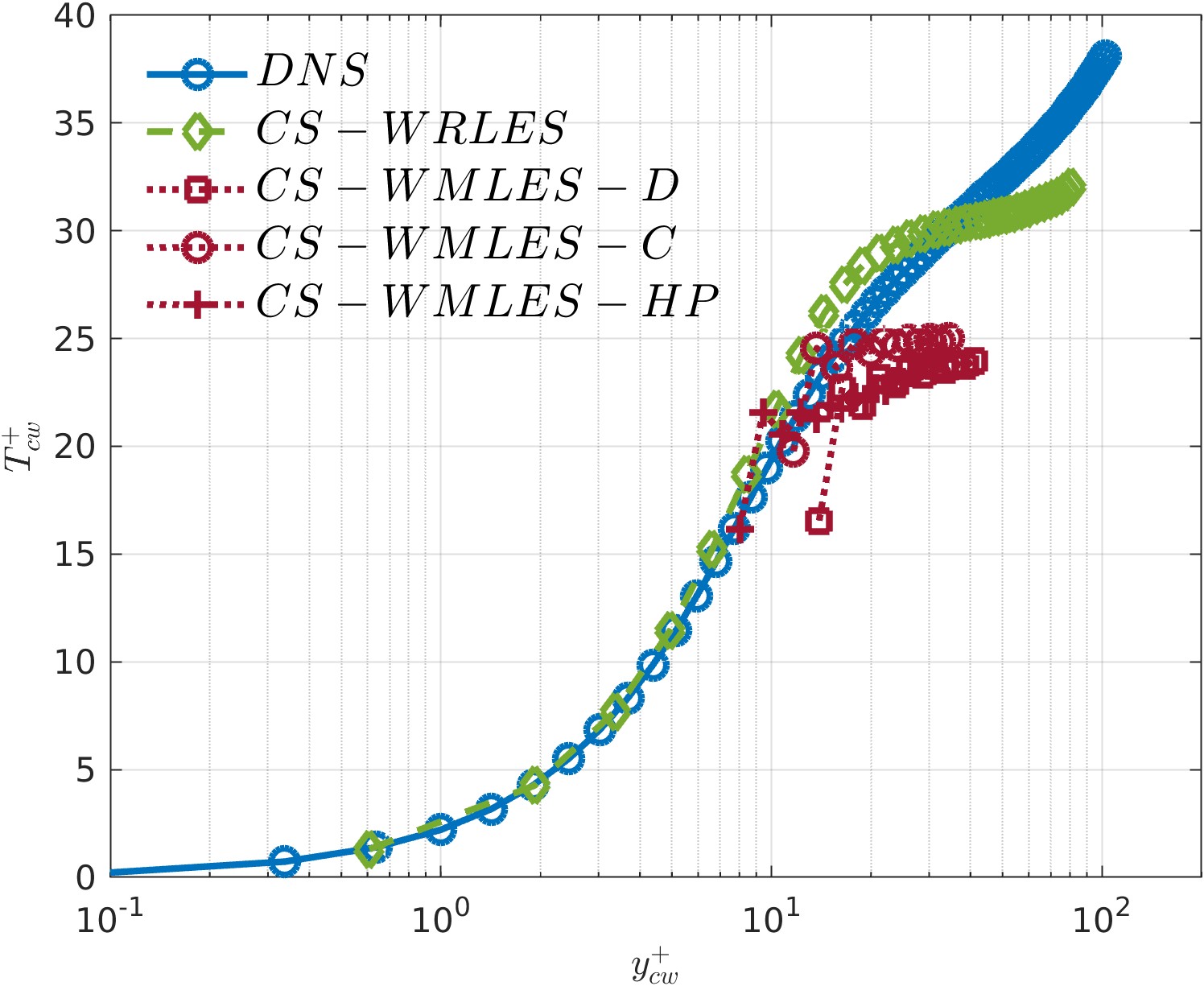}}  \hspace{0.5mm}
    \subfloat[]{\includegraphics[width=0.50\linewidth]{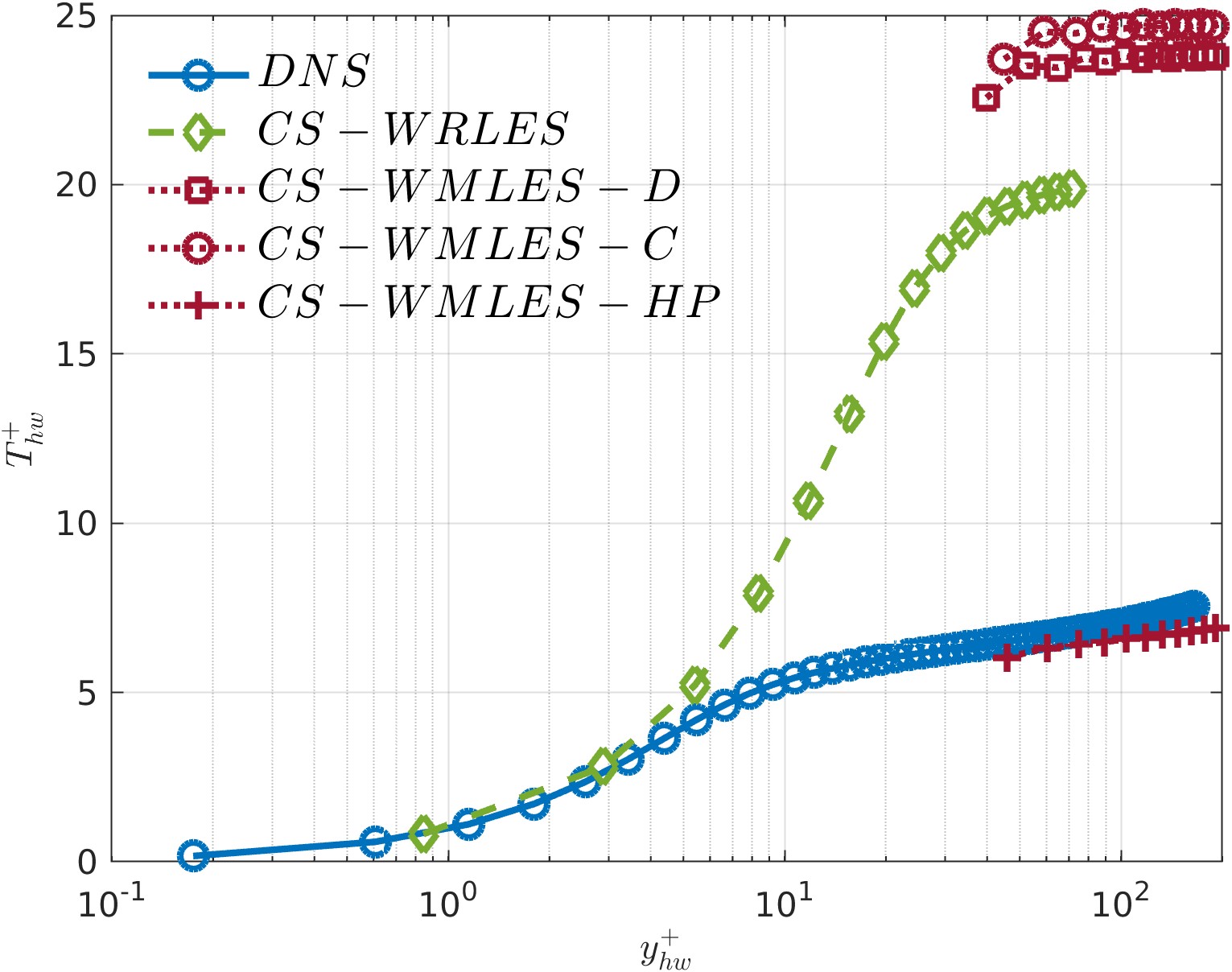}}
	\caption{First-order flow statistics for velocity (a,b) and temperature (c,d) in wall units comparing the WMLES dynamic ``standard law of the wall'' with the CS stress tensor model with respect to the fully-resolved DNS dataset and the CS WRLES for cold (a,c) and hot (b,d) walls.} \label{fig:statistics_WMLES}
\end{figure}

\begin{figure}
	\centering
    \subfloat[]{\includegraphics[width=0.49\linewidth]{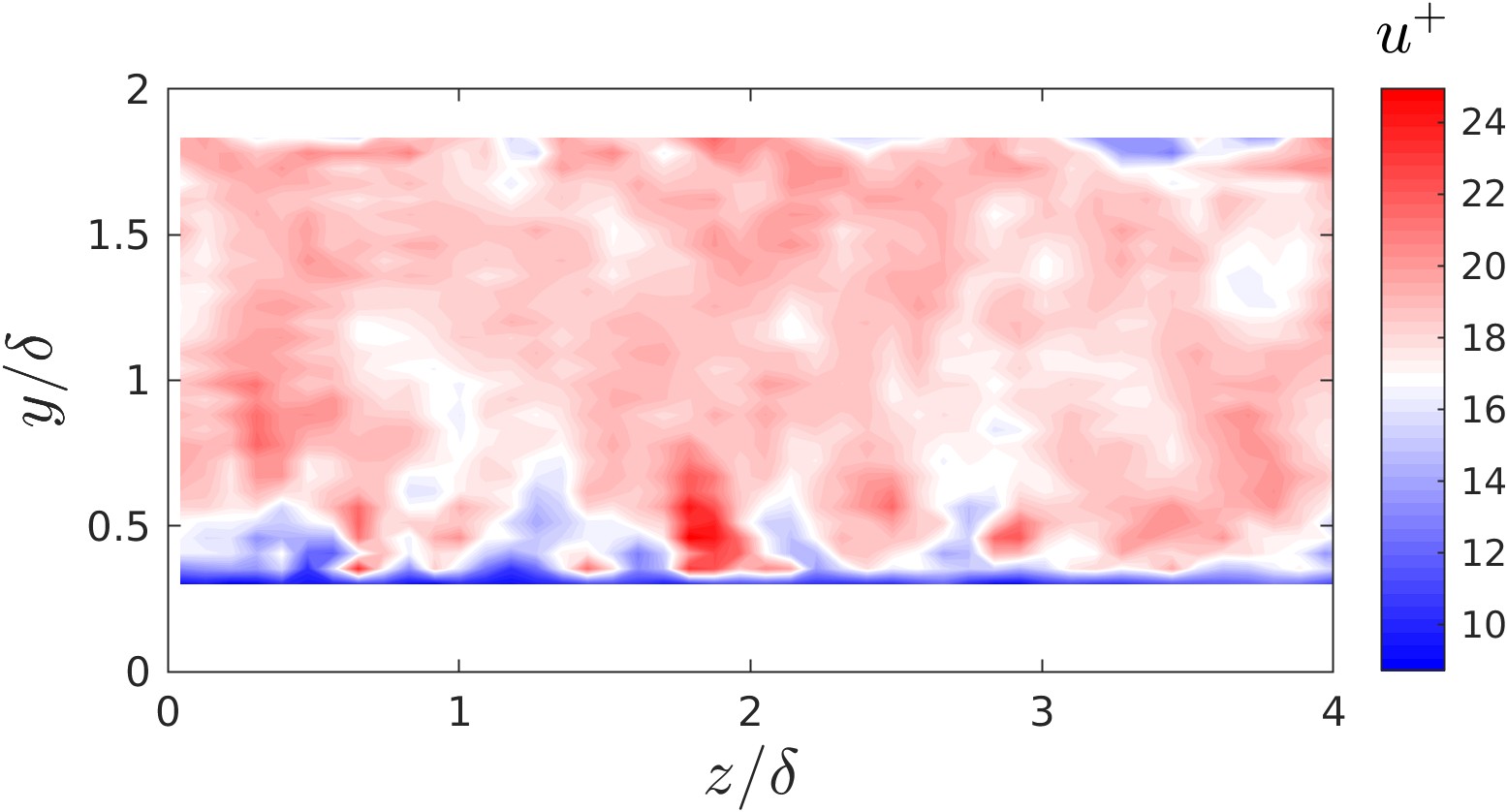}} \hspace{0.5mm}
    \subfloat[]{\includegraphics[width=0.49\linewidth]{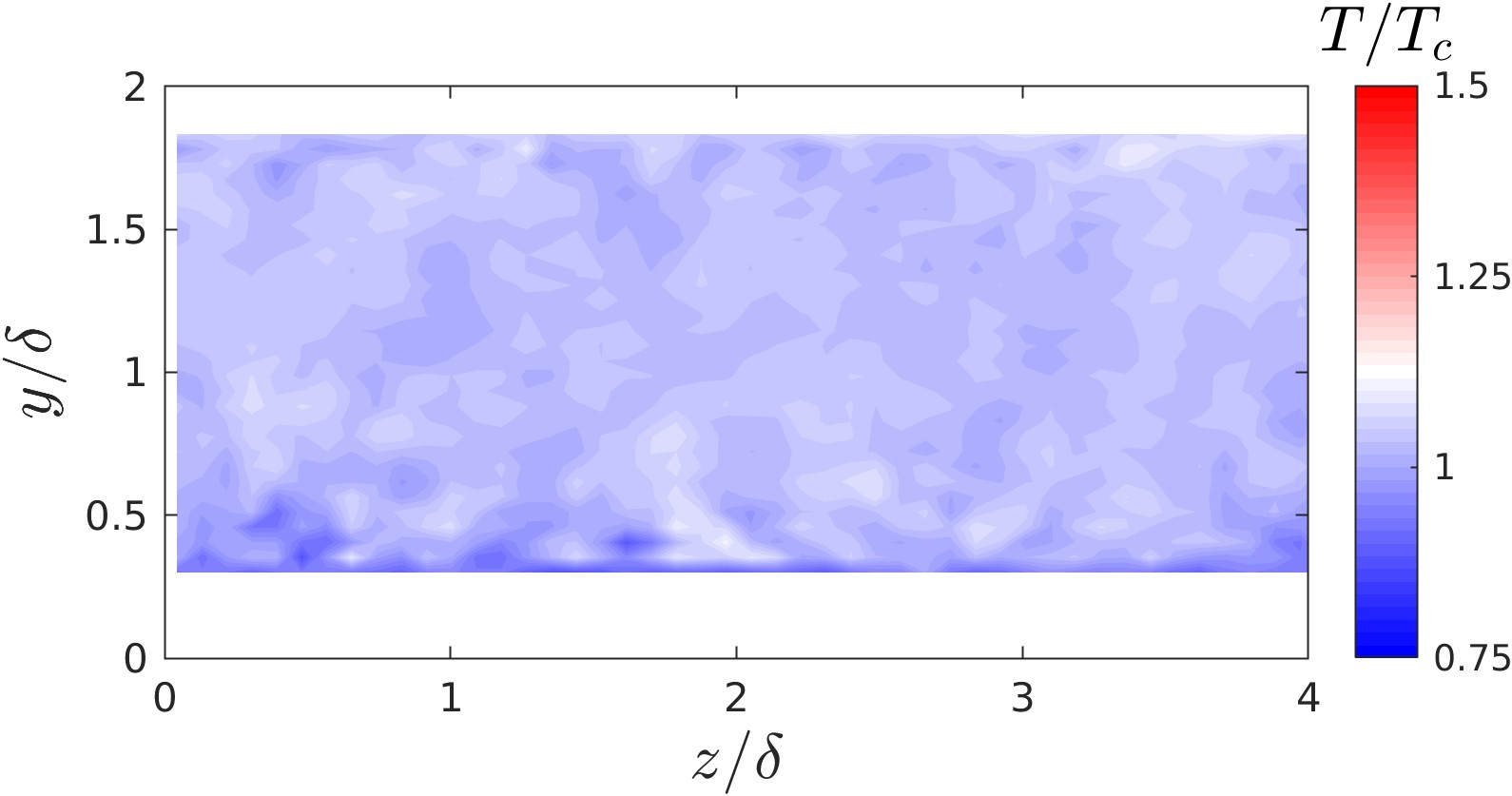}} \\
    \subfloat[]{\includegraphics[width=0.49\linewidth]{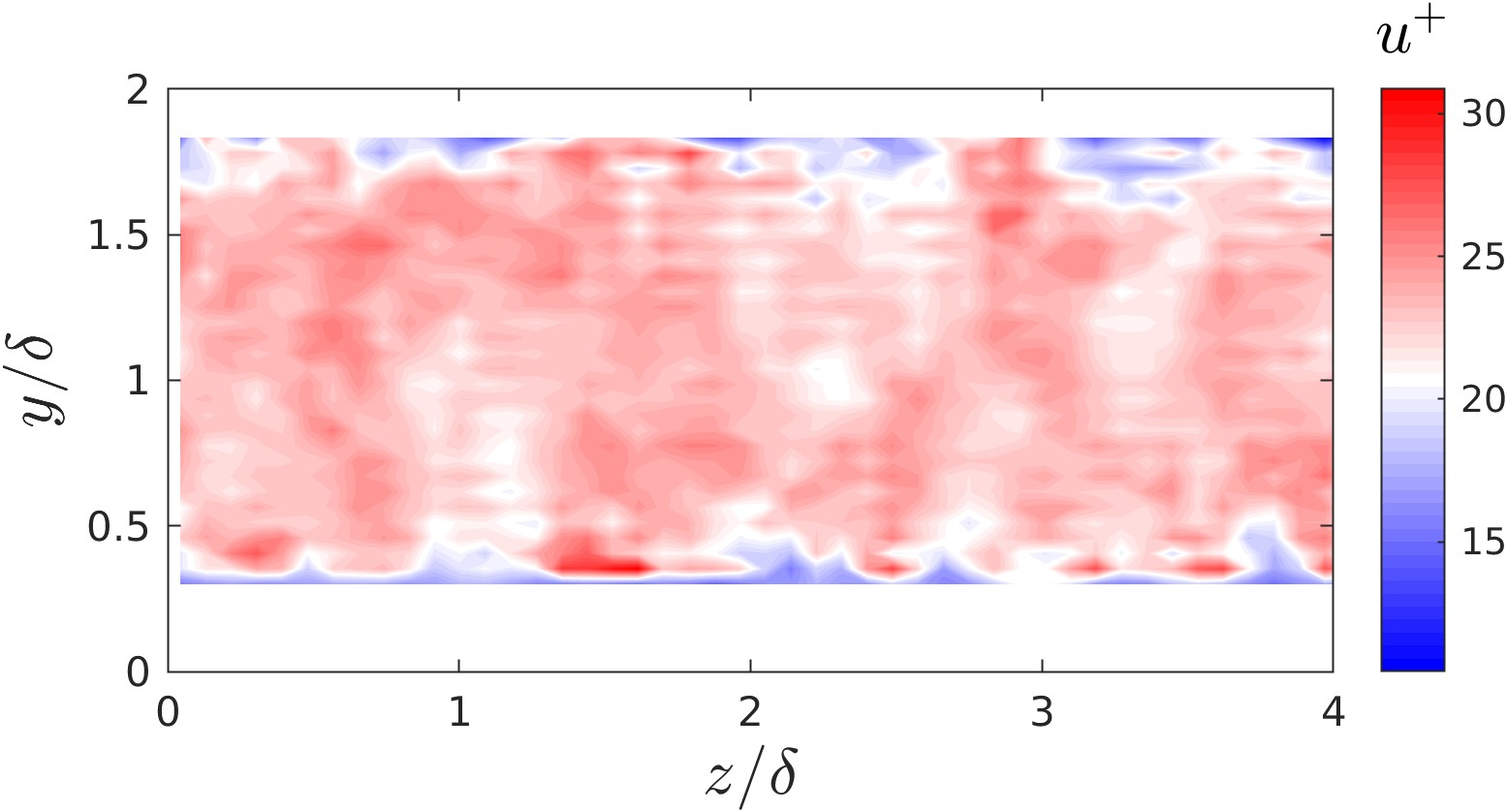}} \hspace{0.5mm}
    \subfloat[]{\includegraphics[width=0.49\linewidth]{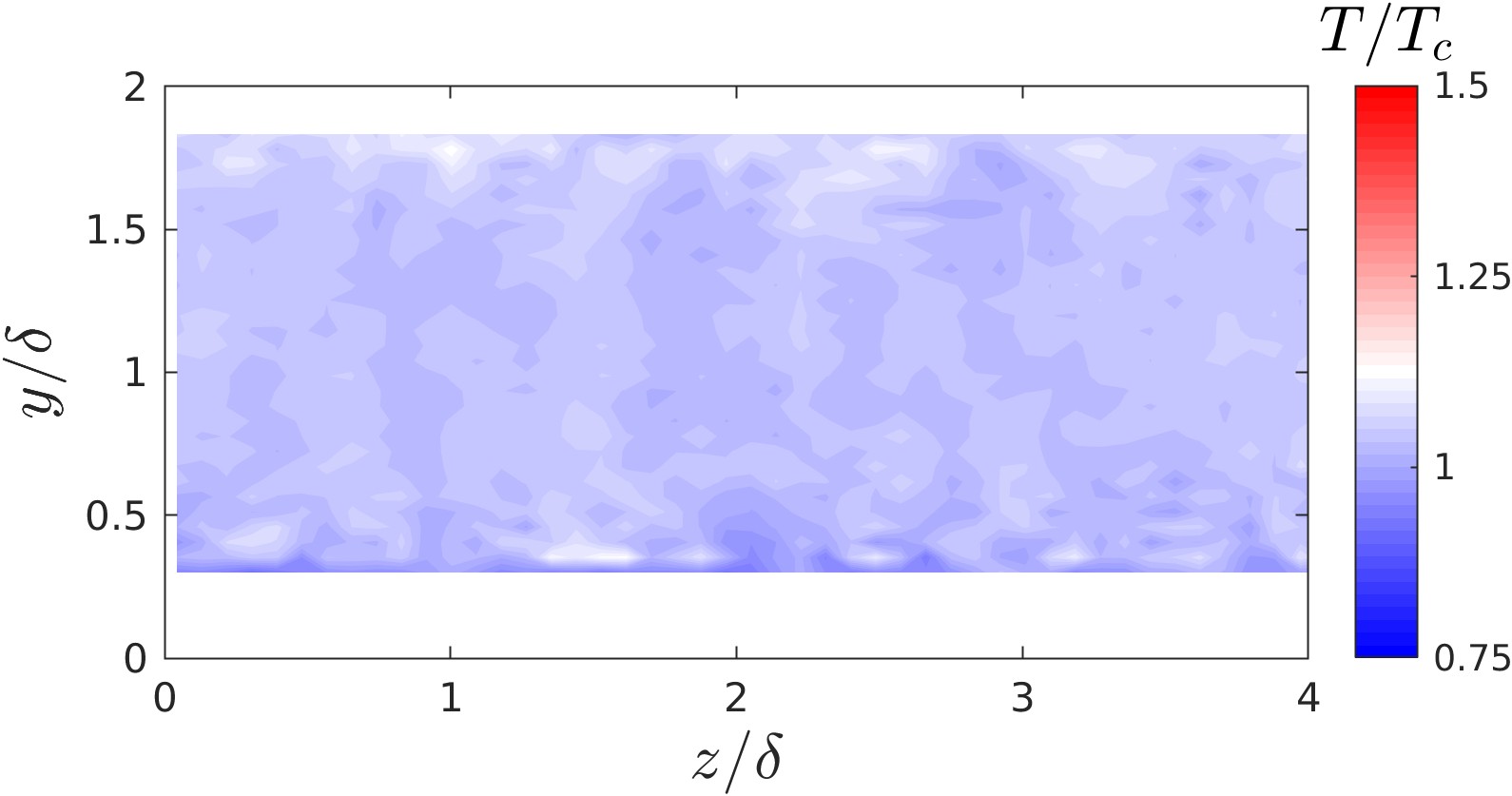}} \\
    \subfloat[]{\includegraphics[width=0.49\linewidth]{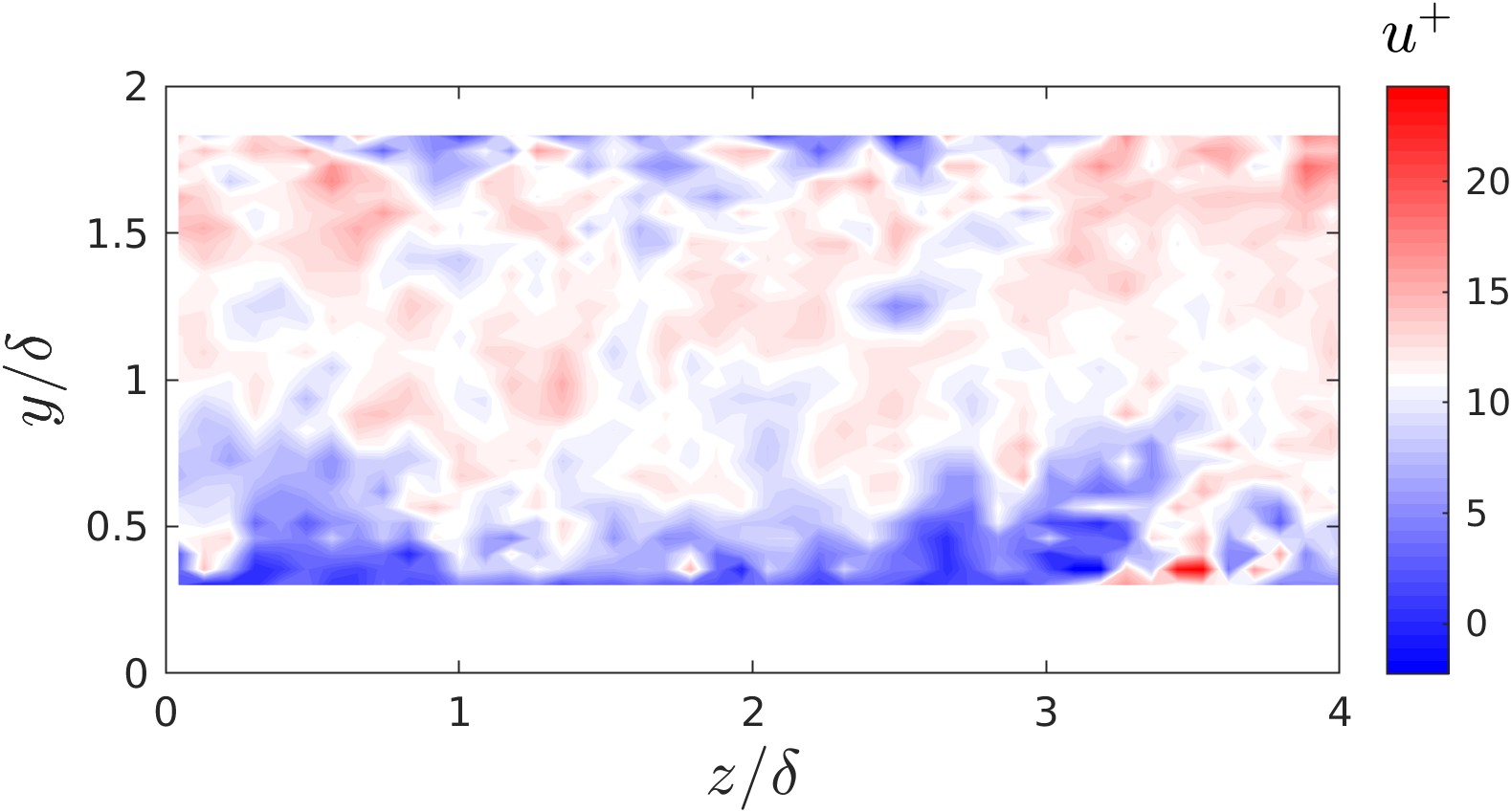}} \hspace{0.5mm}
    \subfloat[]{\includegraphics[width=0.49\linewidth]{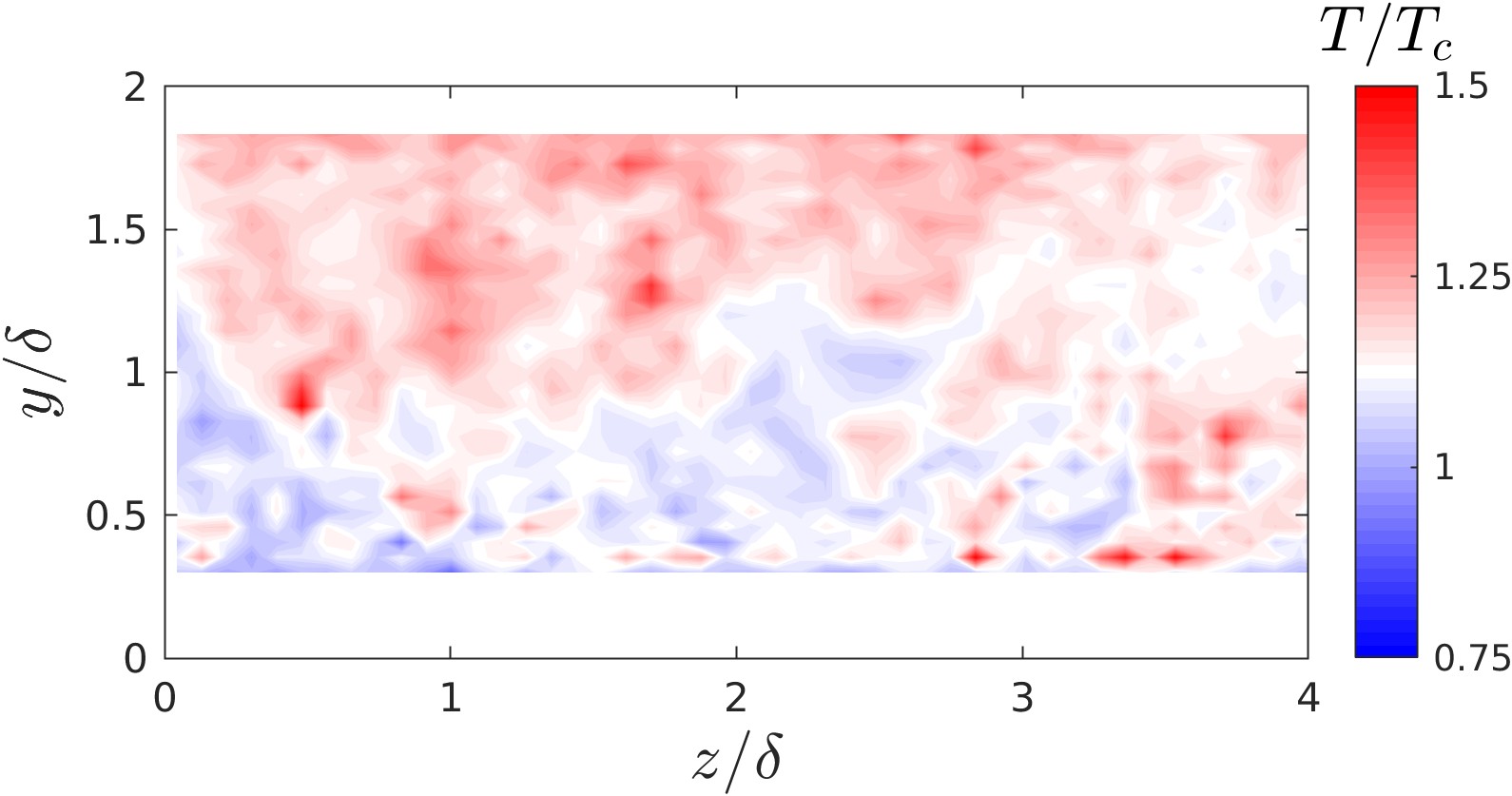}} \\
	\caption{Snapshots along the spanwise and wall-normal directions of velocity in wall units (a,c,e) and temperature normalized by critical point (b,d,f) of the CS WMLES for dynamic (a,b), coupled (c,d) and correlation-based (HP) model (e,f).} \label{fig:snapshots_spanwise_DNS_WMLES_CS}
\end{figure}

The first-order flow statistics for the WMLES cases are depicted in Figure~\ref{fig:statistics_WMLES} in terms of velocity and temperature profiles.
It can be seen from the plots that neither WMLES-D nor WMLES-C accurately capture the behaviour of the velocity and temperature DNS profiles when coupled with the CS SGS stress tensor model.
First, in terms of velocity, WMLES-D can accurately estimate the wall-momentum flux such that the velocity of the first inner point is recovered for the cold wall.
However, it fails to properly represent the rest of the profile.
This highlights the lack of suitability of these SGS stress tensor models for such regimes.
Instead, the WMLES-C does not match the cold wall velocity, but so it does with the temperature.
Moreover, both models significantly overestimate the temperature profile at the hot wall.
In fact, driven by the relatively low accuracy of both WMLES models with respect to the DNS result, a correlation-based wall model, labelled as WMLES-HP, has been proposed.
This method has been derived by optimizing the coefficients of an equation for velocity and temperature with similar structure as the empirical correlations of Eqs.~\ref{eq:u_plus_empirique}-\ref{eq:T_plus_empirique} (details are found in Appendix~\ref{sec:Appendix_C}).
Such model enhances the recovery of the resolved part of velocity and temperature profiles. Nevertheless, it still suffers from inaccurate results due to the effects of the SGS stress tensor model.
Finally, the instantaneous snapshots depicted in Figure~\ref{fig:snapshots_spanwise_DNS_WMLES_CS} demonstrate that the recovery of the first inner point in the vicinity of the pseudo-boiling line region results in a solution similar to the DNS.
The fluctuations seen on the velocity fields are, in this regard, similar to the ones of the WRLES.
However, it is observed that there are prominent signs of better capturing the fluctuations compared to the WRLES approach.

\section{Heat transfer phenomena} \label{sec:heat_transfer}

Table~\ref{tab:dimensionless_comparison} summarizes the resulting dimensionless numbers of heat-transfer phenomena from the WRLES and WMLES computations.
Several observations can be extracted from this table.
First, it is noted that the WRLES cases obtain the same bulk velocity as the DNS, at detriment of wall shear stress and friction velocities.
Instead, the WMLES volumetric force is driven to obtain, analogously as the DNS, the desired shear-stress at walls, yielding, in this case, larger bulk quantities.
Second, as reported in the first-order flow statistics above, only WMLES-HP estimates the friction temperature value similar to the DNS.
Third, the skin-friction coefficient is reasonable well captured for the WMLES at the hot wall, specially WMLES-C, whereas both WRLES and WMLES strategies largely underestimate it at the cold wall by a factor of $2\times$.
Similarly, the WRLES captures correctly the Nusselt number at the cold wall, but not for the hot wall.
Instead, WMLES-D and WMLES-C fairly approximate the $Nu$ at the hot wall.
Nevertheless, \review{they overpredict} by $2\times$ the expected value at the cold wall.
Finally, the Stanton number, which represents the ratio of heat transfer coefficient to heat capacity of the fluid, is only recovered for WMLES-HP at the hot wall.
Nonetheless, all the strategies effectively predict it at the cold wall, \review{specially WRLES.}
It is important to note that the Prandtl number depends on the thermodynamic state confined by the wall temperatures.
Hence, WRLES methods match correctly the DNS values, i.e., same Dirichlet boundary conditions are applied.
However, WMLES strategies slightly deviate as the flux is imposed at the wall in lieu of the target quantity.
In particular, $Pr$ at both wall is overestimated by factors of ${1.5-2}\times$ for WMLES-D and WMLES-C.
Instead, as detailed in Table~\ref{tab:heat_flux_wall}, the WMLES-HP value is closely predicted at the hot wall, but it provides larger values than expected at the cold wall.

\begin{table}

\resizebox{\textwidth}{!}{
\begin{tabular}{ccc|ccccccccc}
                & & & $u_b$ & $Re_b \cdot 10^{3}$ & $u_\tau$ & $T_\tau$ & $Re_{\tau}$ & $C_f \cdot 10^{-3}$ & $N\!u$ & $St \cdot 10^{-2}$\\
    \hline
    \multirow{2}{*}{DNS} & & Hot  & $2.43$ & $6.15$ &$0.19$ & $7.22$ & $180.0$ & $3.46$ & $22.03$ & $12.6$  \\
                                  &  & Cold &  &  & $0.19$ & $0.98$ & $100.0$ & $20.6$ & $5.60$ & $2.61$ \\
    \hline
    \multirow{6}{*}{WRLES} & \multirow{2}{*}{CS} & Hot & $2.41$ & $5.74$ &$0.15$ & $2.89$ & $144.7$ & $1.97$ & $6.52$ & $4.64$  \\
                                  &  & Cold  &  &  &$0.11$ & $0.93$ & $59.4$ & $6.44$ & $3.67$ & $2.88$ \\
    & \multirow{2}{*}{DS} & Hot & $2.41$ & $5.77$ &$0.14$ & $2.72$ & $137.8$ & $1.89$ & $5.98$ & $4.47$ \\
                                  &  & Cold &  &  &$0.11$ & $0.97$ & $58.9$ & $6.67$ & $3.62$ & $2.86$  \\
    & \multirow{2}{*}{WALE} & Hot & $2.41$ & $5.75$ &$0.15$ & $2.89$ & $141.8$ & $1.97$ & $6.50$ & $4.72$ \\
                                  &  & Cold & & &$0.12$ & $1.00$ & $60.9$ & $7.05$ & $3.92$ & $3.00$ \\ 
    \hline
    \multirow{6}{*}{$\textrm{WMLES}-\textrm{D}$} & \multirow{2}{*}{CS} & Hot & $3.33$ & $8.85$ &$0.22$ & $2.28$ & $258.8$ & $4.76$ & $15.4$ & $3.75$  \\
                                  &  & Cold  &  &  &$0.19$ & $1.37$ & $50.4$ & $10.1$ & $9.12$ & $4.03$ \\
    & \multirow{2}{*}{DS} & Hot & $3.17$ & $8.39$ &$0.22$ & $2.37$ & $253.6$ & $5.22$ & $15.3$ & $3.89$  \\
                                  &  & Cold  &  &  &$0.19$ & $1.44$ & $48.0$ & $11.0$ & $9.75$ & $4.28$ \\
    & \multirow{2}{*}{WALE} & Hot & $3.15$ & $8.34$ &$0.22$ & $2.37$ & $252.0$ & $5.22$ & $15.2$ & $3.88$  \\
                                  &  & Cold  &  &  &$0.19$ & $1.44$ & $48.0$ & $11.2$ & $9.74$ & $4.28$ \\ 
    \hline
    \multirow{6}{*}{$\textrm{WMLES}-\textrm{C}$} & \multirow{2}{*}{CS} & Hot & $4.23$ & $11.6$ &$0.25$ & $2.18$ & $292.5$ & $3.88$ & $16.9$ & $3.66$  \\
                                  &  & Cold  &  &  &$0.20$ & $1.33$ & $42.2$ & $6.39$ & $8.82$ & $3.80$ \\
    & \multirow{2}{*}{DS} & Hot  & $4.11$ & $11.3$ &$0.25$ & $2.24$ & $291.7$ & $4.20 $ & $17.1$ & $3.75$  \\
                                  &  & Cold  &  &  &$0.19$ & $1.38$ & $40.2$ & $6.76$ & $9.27$ & $3.96$ \\
    & \multirow{2}{*}{WALE} & Hot & $4.12$ & $11.3$ &$0.25$ & $2.23$ & $292.5$ & $4.17$ & $17.1$ & $3.74$  \\
                                  &  & Cold  &  &  &$0.20$ & $1.38$ & $40.2$ & $6.73$ & $9.23$ & $3.94$ \\    
    \hline
    \multirow{6}{*}{$\textrm{WMLES}-\textrm{HP}$} & \multirow{2}{*}{CS} & Hot & $3.21$ & $9.38$ &$0.27$ & $5.50$ & $297.7$ & $7.59$ & $47.4$ & $12.2$  \\
                                  &  & Cold  &  &  &$0.19$ & $2.02$ & $29.0$ & $16.5$ & $10.0$ & $4.08$ \\
    & \multirow{2}{*}{DS} & Hot  & $3.09$ & $9.08$ &$0.27$ & $5.57$ & $297.8$ & $8.34$ & $48.8$ & $12.5$  \\
                                  &  & Cold  &  &  &$0.20$ & $2.27$ & $26.3$ & $17.8$ & $11.2$ & $4.53$ \\
    & \multirow{2}{*}{WALE} & Hot & $3.26$ & $9.51$ &$0.29$ & $5.57$ & $312.8$ & $8.28$ & $52.0$ & $12.5$  \\
                                  &  & Cold  &  &  &$0.20$ & $2.02$ & $28.2$ & $16.9$ & $10.7$ & $4.28$ \\  
    \hline
  \end{tabular}}

  \caption{Dimensionless numbers obtained from the WRLES and WMLES cases at the cold and hot walls; the units of velocity-based values correspond to $m/s$.} \label{tab:dimensionless_comparison}
  
\end{table}

\begin{figure}
	\centering
	\subfloat[]{\includegraphics[width=0.49\linewidth]{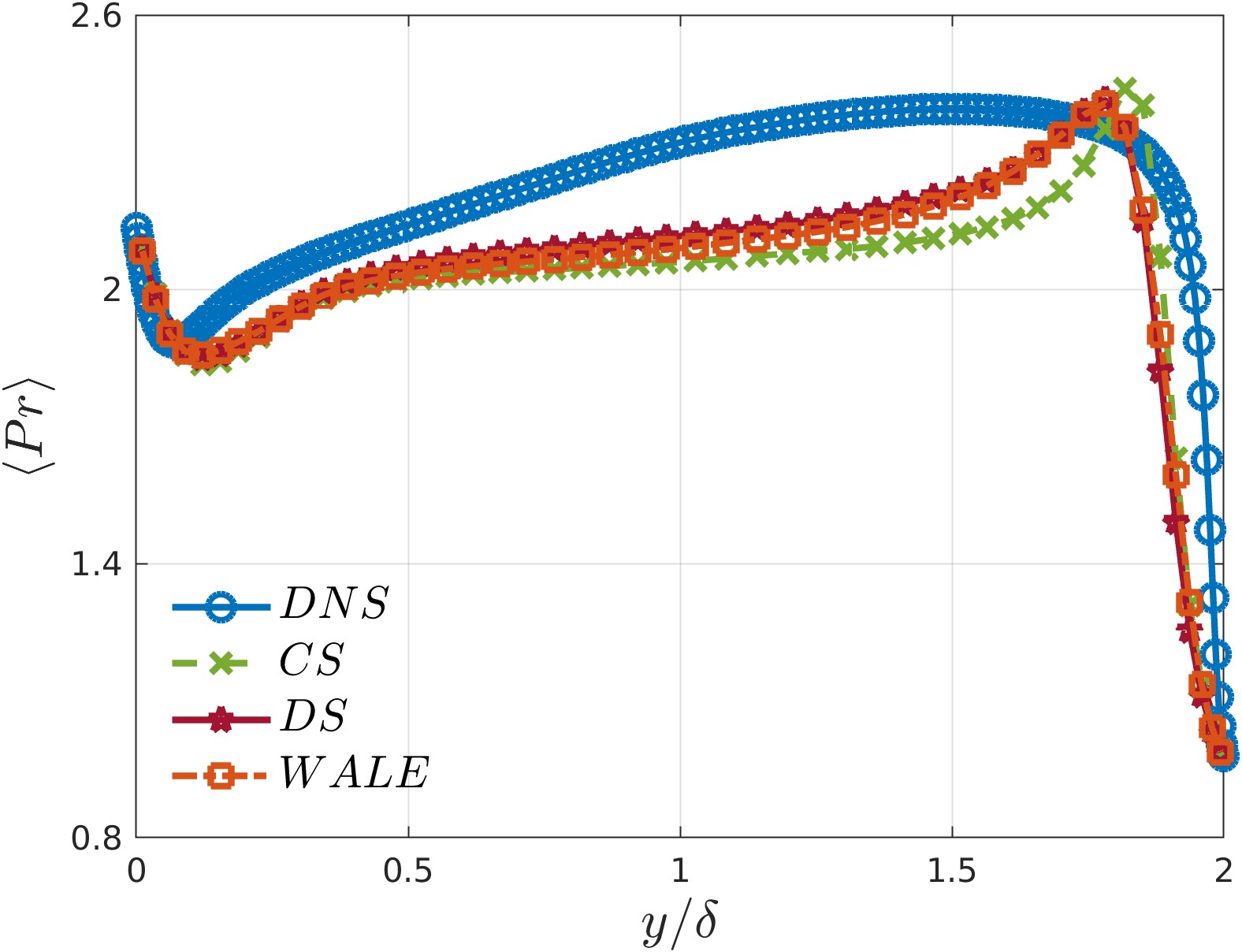}} \hspace{0.5mm}
    \subfloat[]{\includegraphics[width=0.49\linewidth]{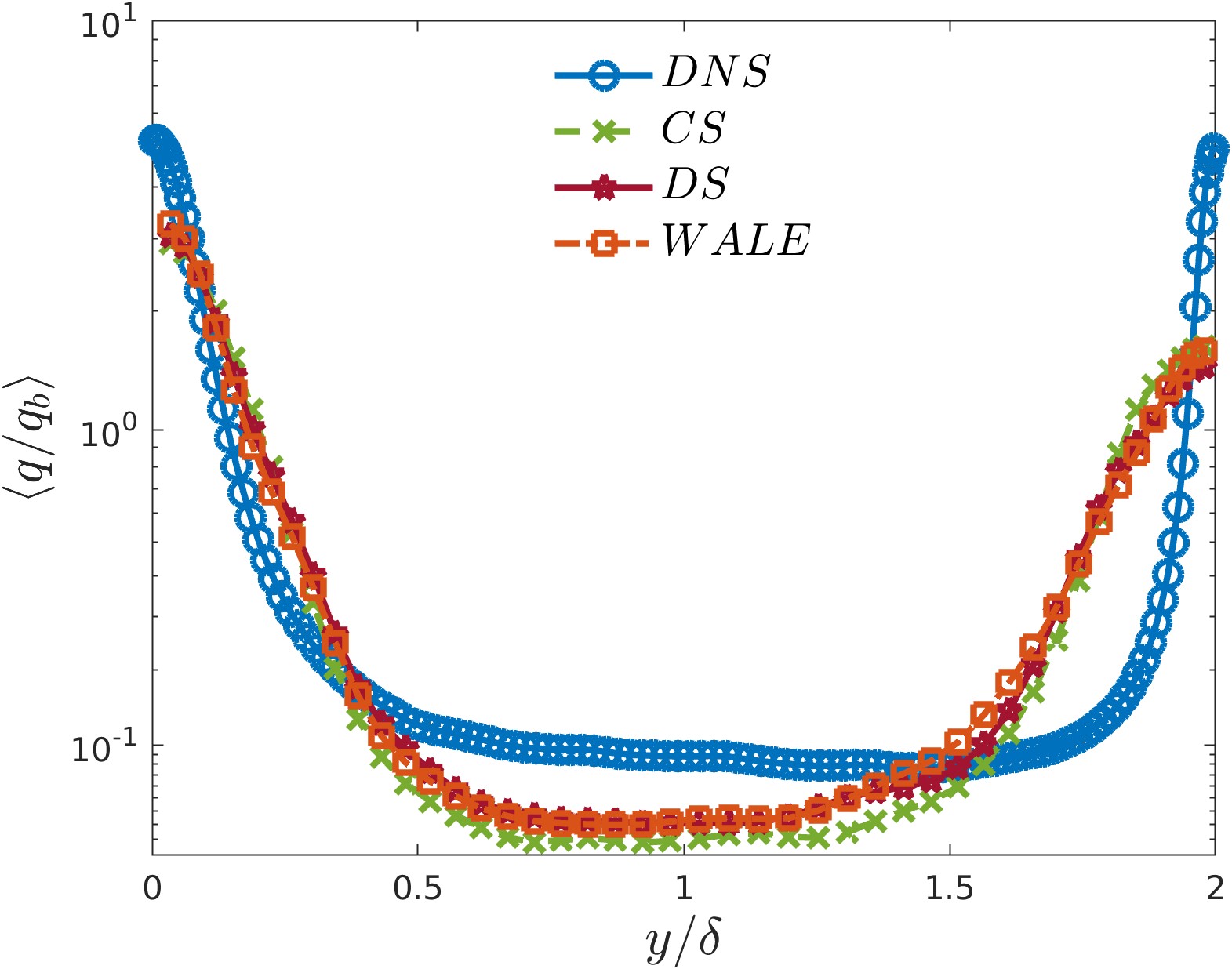}} \\
	\caption{Prandtl number (a) and heat flux (b) along the wall-normal direction normalized by the channel half-height ($y/\delta$) comparing WRLES SGS models against DNS. Heat flux is normalized by bulk heat flux of value $q_b = 6.12 \cdot 10^4$ and $q_b = 6.00 \cdot 10^4 \thinspace W/m^2$ for DNS and WRLES, respectively.} \label{fig:Pr_q_vs_y_WRLES}
\end{figure}

\begin{figure}
	\centering
	\subfloat[]{\includegraphics[width=0.49\linewidth]{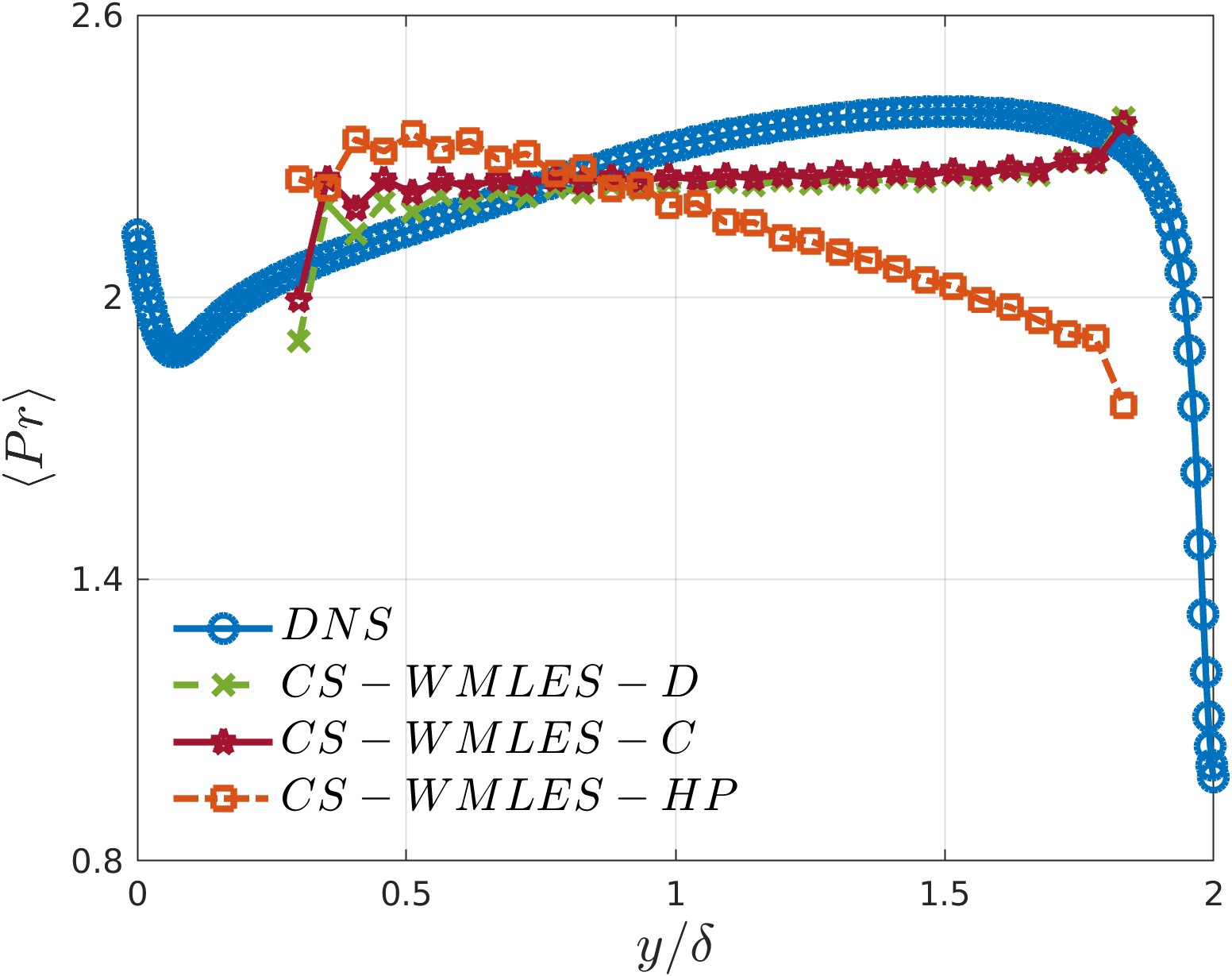}} \hspace{0.5mm}
    \subfloat[]{\includegraphics[width=0.49\linewidth]{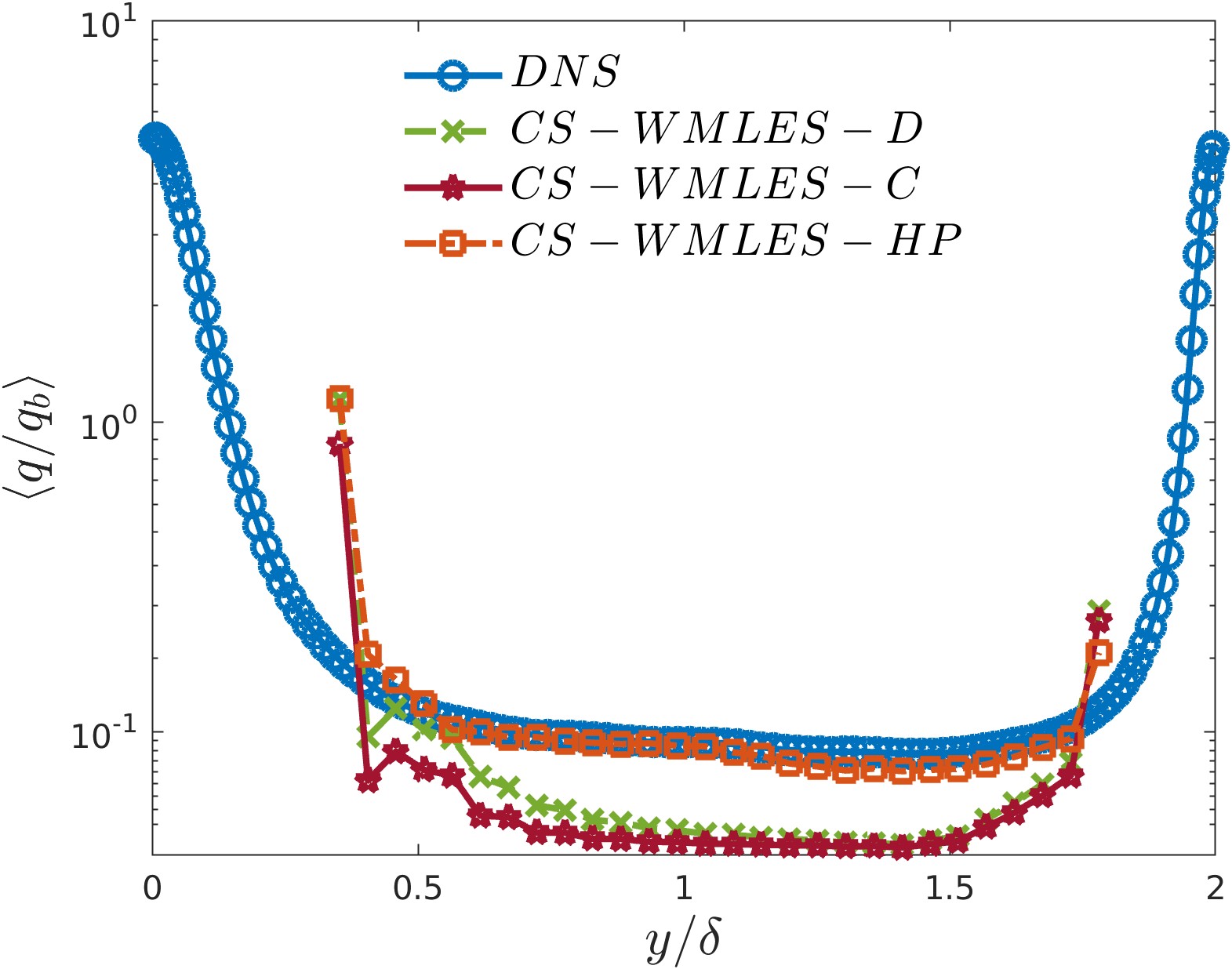}} \\
	\caption{Prandtl number (a) and heat flux (b) along the wall-normal direction normalized by the channel half-height ($y/\delta$) comparing CS WMLES models against DNS. Heat flux is normalized by bulk heat flux of value $q_b = 6.12 \cdot 10^4$, $3.88 \cdot 10^4$, $3.39 \cdot 10^4$ and $3.16 \cdot 10^4$ \thinspace $W/m^2$ for DNS, WMLES-D, WMLES-C and WMLES-HP, respectively.} \label{fig:Pr_q_vs_y_WMLES}
\end{figure}

Figure~\ref{fig:Pr_q_vs_y_WRLES} shows the time-averaged profile of Prandtl number and heat flux along the wall-normal direction for the WRLES computations.
Several observations can be extracted from the plots.
First, the Prandtl number evolves from a value of $Pr \approx 2.2$ at the cold wall to a value of $Pr \approx 0.9$ at the hot wall.
Thus, the importance of the momentum and thermal boundary layers changes between the cold and hot walls.
The SGS models capture well the trend and also the wall values. In addition, they successfully capture the oscillatory behaviour in the vicinity of the cold wall.
Second, the heat fluxes are in general well predicted by the WRLES cases.
Analogously, Figure~\ref{fig:Pr_q_vs_y_WMLES} represents the same quantities utilizing the CS model for the WMLES strategies.
In terms of Prandtl number, they fail to correctly represent the trend along the y-direction, presenting a flat behaviour for WMLES-D and WMLES-C, worsening for the WMLES-HP case.
Finally, the heat flux at the center of the channel is similar to the one obtained by the DNS behaviour.
In particular, the wall heat-fluxes are better approximated by the WMLES strategies as quantified in Table~\ref{tab:heat_flux_wall}.

\begin{table}

\resizebox{\textwidth}{!}{
\begin{tabular}{ccc|cccc}
                & & & $DNS$ & $WMLES-D$ & $WMLES-C$ & $WMLES-HP$\\
    \hline
    \multirow{2}{*}{$q_w \cdot 10^{5} \thinspace (W/m^2)$} & & Hot  & $3.05$ & $3.50$ & $3.80$ & $5.72$ \\
                                        & & Cold & $3.14$ & $4.76$ & $4.77$ & $7.77$ \\
    \hline
    \multirow{2}{*}{$Pr$} & & Hot  & $0.97$ & $1.58$ & $1.58$ & $1.30$ \\
                                        & & Cold & $2.14$ & $4.49$ & $5.51$ & $8.47$ \\
    \hline
                                      
  \end{tabular}}

  \caption{Heat flux and Prandtl number at the cold and hot walls for DNS and WMLES with CS.} \label{tab:heat_flux_wall}
  
\end{table}

\section{Conclusions}\label{sec:conclusions}

The framework proposed by recent a \textit{a priori} analysis has been implemented and executed with existing SGS stress tensor models and closure expressions for the terms whose activity was found to be important.
The results have proved the suitability of this framework for wall-bounded high-pressure transcritical turbulence.
However, research efforts are still needed in terms of turbulence modeling, particularly, at relatively low-Reynolds-numbers.
The regime explored along with the strong non-linearities encountered when operating across the pseudo-boiling line in non-isothermal conditions exacerbate the difficulties of the SGS models to correctly obtain the velocity and temperature profiles in such flows.
Consequently, both wall-resolved and wall-modeled approaches suffer to recover the fully-resolved DNS results. The differences are mainly driven by the SGS models utilized.
As also concluded in the \textit{a priori} analysis, specific SGS developments are still needed for such type of flows.
Moreover, the dynamic and coupled wall models explored significantly enhance the recovery of the first-order flow statistics.
In particular, the first inner point in the log-law region is well characterized, but the rest of the profile still presents deviations with respect to the DNS result.
This is mainly due to the SGS stress models of the momentum equation.

In terms of wall values, the WRLES approach can fairly estimate the Nusselt number at the cold wall and skin-friction coefficient at the hot wall by \review{$70\%$}.
In addition, the trends along the wall-normal direction of Prandtl number and heat flux are relatively well captured.
In detail, the shape and minimum and maximum values of Prandtl number are also relatively well captured.
Nevertheless, larger differences for the heat flux are observed near the pseudo-boiling region.
The WMLES strategies noticeable improve the profiles of first-order flow statistics.
In particular, the models successfully predict the quantities in the first inner point, but not at the center of the channel as both WRLES and WMLES models overpredict the velocity statistics.
The dynamic model is the preferred option for velocity, specially for the cold wall, and so is the coupled for the temperature estimation.
Nonetheless, they both fail to recover the temperature profiles at the hot wall due to the strong thermophysical gradients in the pseudo-boiling region driving this regime.
To this extent, a correlation-based method was proposed to (i) properly recover the velocity and temperature first inner point resolved quantities, and (ii) predict the velocity and temperature profile at the log-law region of the hot wall, which is altered by pseudo-boiling effects.
The performance of the various SGS stress tensor models assessed is similar.
In particular, the CS is recommended for WMLES-D and DS/WALE for WMLES-C.
This combination also results in a better heat transfer phenomena performance with respect to the DNS results.
Regarding the dimensionless numbers at walls, WMLES-D and WMLES-C recover the Nusselt number at the hot wall, although the cold wall is overestimated by $2\times$.
Furthermore, the skin-friction coefficient is well-captured at the hot wall, but not at the cold wall.
To this end, WMLES heat transfer performance fails to attain the Prandtl profile, and overestimates its wall quantities. Instead, heat flux at wall is faithfully obtained by WMLES-D and WMLES-C, while its trend is quite precisely achieved by WMLES-HP.
In fact, the proposed WMLES-HP yields the expected Stanton number at the hot wall suggesting the suitability to obtain the correct ratio of heat transfer coefficient to fluid heat capacity.

The low-Reynolds-number regime considered in this work stretches the turbulence modeling complexity, i.e, near-wall viscous fluxes are still important and the wall model domain is confined far from the wall.
Therefore, potential enhancement, with this SGS models, may be reflected at larger Reynolds regimes.
Similarly, it is known that smoother thermodynamic non-linear conditions can be well predicted by the coupled wall model.
Nevertheless, as recommended in the \textit{a priori} work, further investigation is needed toward the development of SGS stress tensor models suitable for wall-bounded high-pressure transcritical turbulent flow applications.
Moreover, physics-based models for the unclosed terms of the pressure transport equation are essential to recover the heat transfer fluxes across the domain.
In addition, the ILA model needs to be extended with different filters, or supplementary models, to properly close the SGS of the equation of state.
Finally, exploration of more additional WMLES strategies, which could deal with stronger thermophysical variations, needs to be considered in future work.

\backmatter




\section*{Acknowledgments}

The authors gratefully acknowledge the \textit{Formaci\'o de Professorat Universitari} scholarship (FPU-UPC R.D 103/2019) of the Universitat Polit\`ecnica de Catalunya $\cdot$ BarcelonaTech (UPC) (Spain), the \textit{Serra H\'unter} and SGR (2021-SGR-01045) programs of the Generalitat de Catalunya (Spain), and the computer resources at FinisTerrae III \& MareNostrum and the technical support provided by CESGA \& Barcelona Supercomputing Center (RES-IM-2023-1-0005, RES-IM-2023-2-0005). \textcolor{red}{The financial support for M. Bernades from the ERASMUS+ traineeship and international mention grant is gratefully acknowledge.}


\section*{Compliance with Ethical Standards}

\noindent \textbf{Funding:} Not applicable.

\noindent\textbf{Conflict of Interest:} The authors declare that they have no conflict of interest.

\noindent \textbf{Ethical Approval:} This article does not contain any studies with human participants or animals performed by any of the authors.

\noindent \textbf{Informed consent:} Not applicable.

\noindent \textbf{Authors' contributions:}
Marc Bernades: Conceptualization, Formal analysis, Investigation, Software, Writing – original draft; 
Florent Duchaine: Conceptualization, Investigation, Writing – review, editing; 
Francesco Capuano: Conceptualization, Investigation, Writing – review, editing; 
Lluís Jofre: Conceptualization, Investigation, Writing – review, editing.

\noindent \textbf{Data Availability Statement:} At request.

\begin{appendices}






\section{Low-pressure isothermal turbulent channel flow}\label{sec:Appendix_A}

\begin{figure}
	\centering
	{\includegraphics[width=0.60\linewidth]{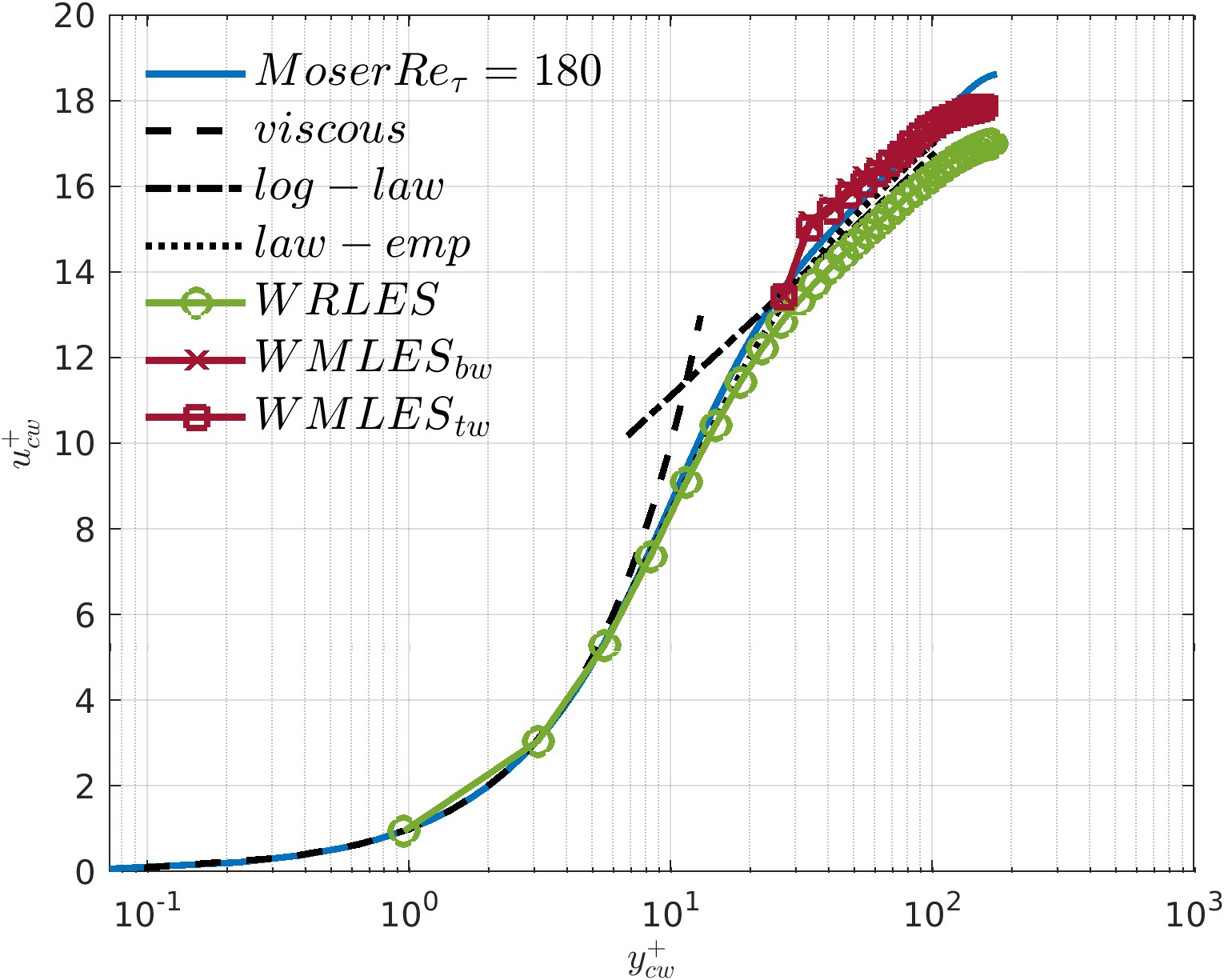}}
	\caption{First-order statistics for velocity in wall units comparing the reference solution with WRLES and WMLES and the ``standard law of the wall''.} \label{fig:statistics_u_LP}
\end{figure}

\begin{figure}
 \centering
 \subfloat[]{\includegraphics[width=0.9\linewidth]{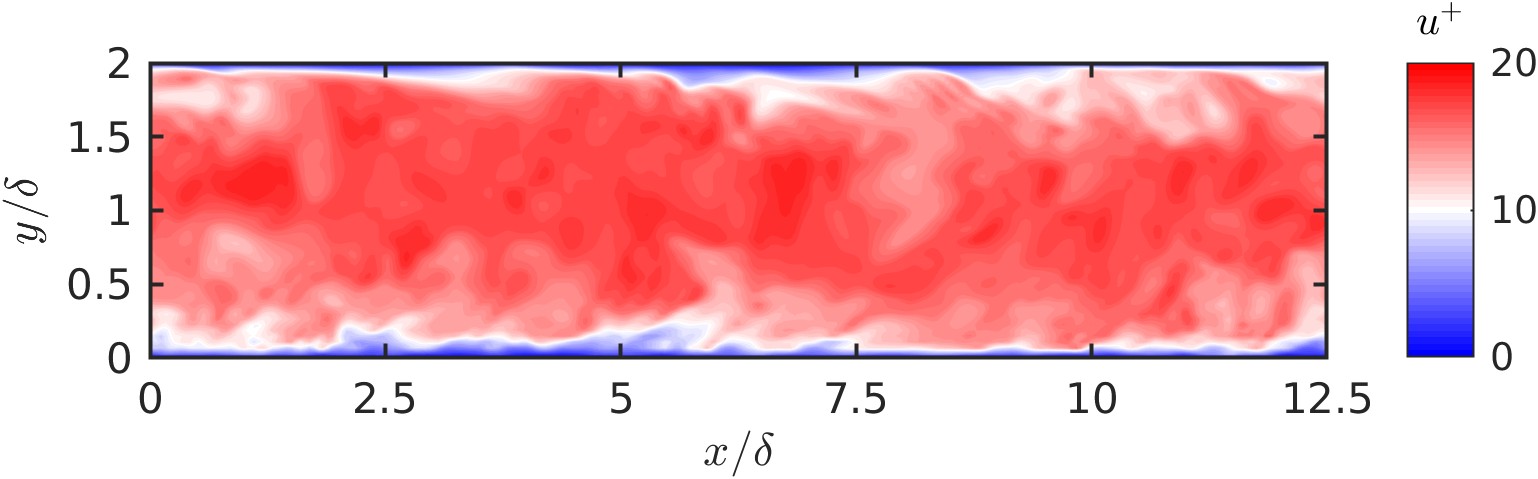}}
\\
 \subfloat[]{\includegraphics[width=0.9\linewidth]{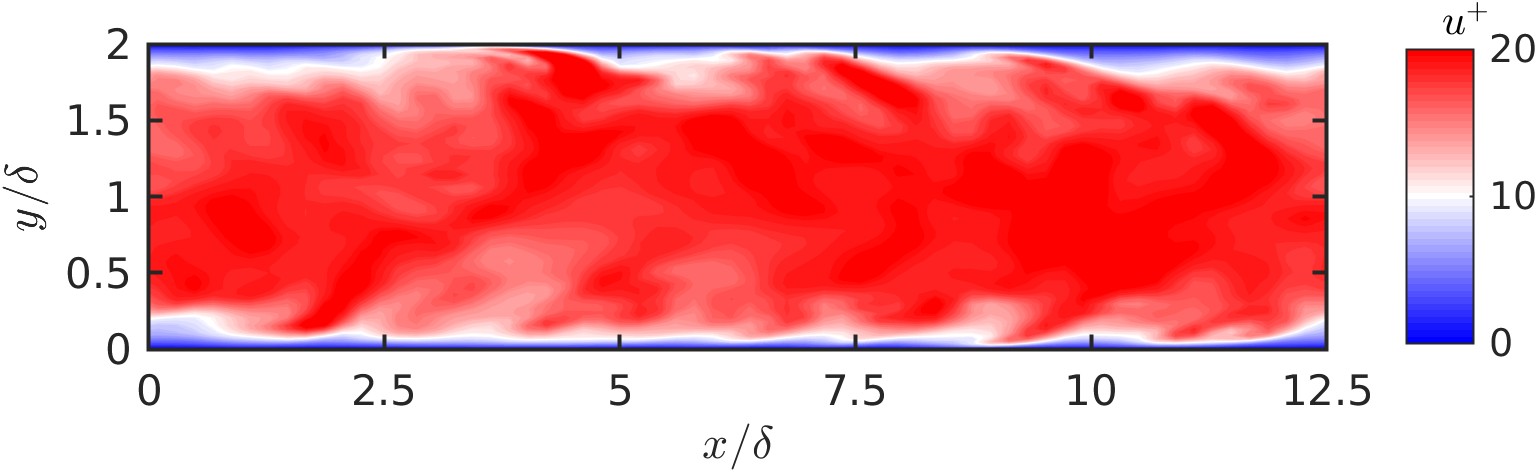}}
 \\
 \subfloat[]{\includegraphics[width=0.9\linewidth]{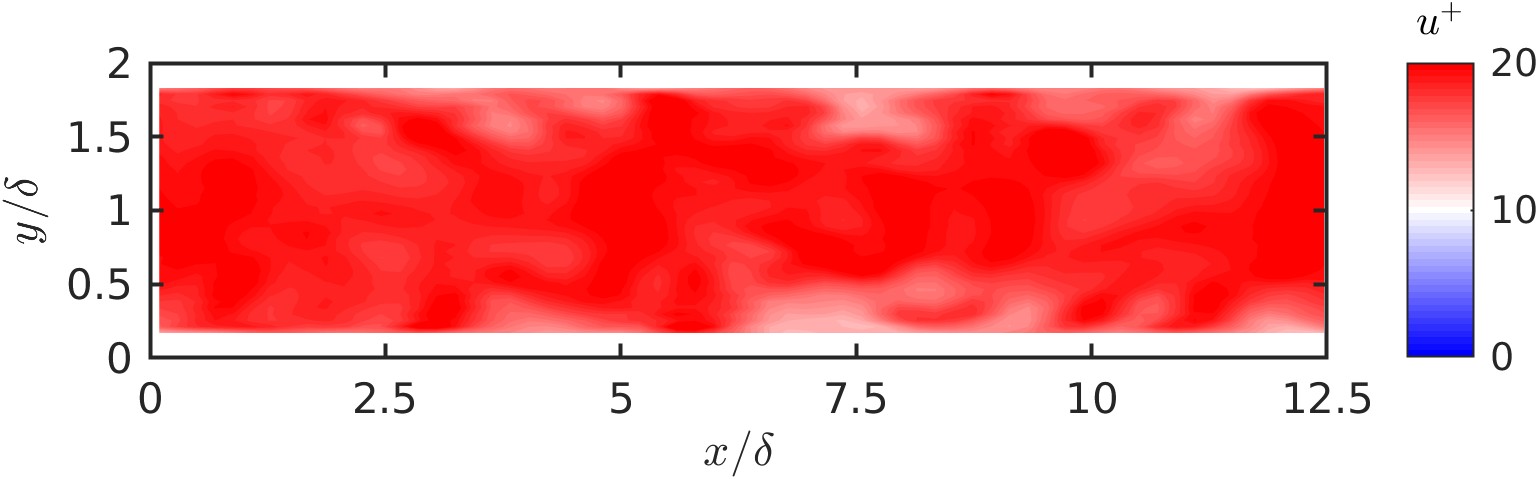}}
\vspace{0mm}
 \caption{Snapshot of instantaneous streamwise velocity in wall units $u^+$ (cw) on a $x$-$y$ slice for (a) DNS, (b) WRLES and (c) WMLES.}  \label{fig:3d_turbulent_channel_flow_contours}
\end{figure}

The WRLES and WMLES implementations have been firstly validated against the isothermal channel flow at low-pressure (ideal-gas) conditions with $Re_\tau = 180$~\cite{Moser1999-A}.
Therefore, based on this setup, an equivalent DNS has been computed with in-house flow solver~\cite{RHEA2023-A,Abdellatif2023-A} used for this work at similar grid resolutions. Consequently, the WRLES was computed with the Classical Smagorinsky SGS stress tensor model with the first grid point at $y^+ = 1$ (also with WALE which, for brevity, is not shown in the results). Finally, the WMLES has been implemented with the ``standard law of the wall'' and with the same SGS stress tensor models. In this case, the first grid point was set at $y^+ = 30$ with uniform grid in the remaining inner domain. This domain resolution is set to be similar to the WRLES center channel.
In detail, the selected mesh sizes and the corresponding resolutions for each case are summarized in Table~\ref{tab:resolutions_LP_channelflow} (note resolution are numerically computed and small differences are captured for same mesh size).
To this end, Figure~\ref{fig:statistics_u_LP} presents the first-order flow statistics. Clearly, both WRLES and WMLES results correctly recover the statistics, and therefore their implementation and the solver is validated.
In addition, Figure~\ref{fig:3d_turbulent_channel_flow_contours} depicts the instantaneous fields of normalized velocity in inner scales for DNS, WRLES and WMLES.

\begin{table}

\begin{tabular}{c|cccc}
     Case & $N_x \times N_y \times N_z $ & $\Delta x^+$ & $\Delta y^+$ & $\Delta z^+$\\
    \hline
    {\citet{Moser1999-A}}   & $128\times129\times128$ & $17.7$ & $0.3 - 4.4$ & $5.9$ \\
    DNS    & $256\times128\times128$ & $9.5$ & $0.1 - 4.0$ & $6.3$ \\
    WRLES  & $64\times64\times64$    & $35.6$ & $1.0 - 7.1$ & $11.9$ \\
    WMLES  & $64\times40\times64$    & $33.9$ & $6.9 - 30.0$ & $11.3$ \\
    
    \hline
                                   
  \end{tabular}

  \caption{Mesh resolution in wall units for the low-pressure isothermal turbulent channel flow at $Re_\tau = 180$.} \label{tab:resolutions_LP_channelflow}
  
\end{table}

\section{Activity of the SGS terms of the pressure transport equation}\label{sec:Appendix_B}

\begin{figure}
	\centering
	\subfloat[]{\includegraphics[width=0.485\linewidth]{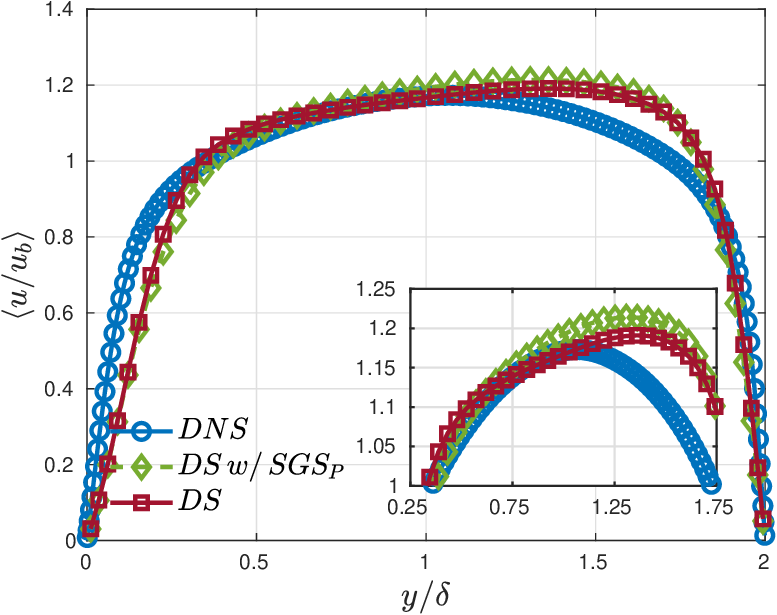}} \hspace{0.5mm}
    \subfloat[]{\includegraphics[width=0.495\linewidth]{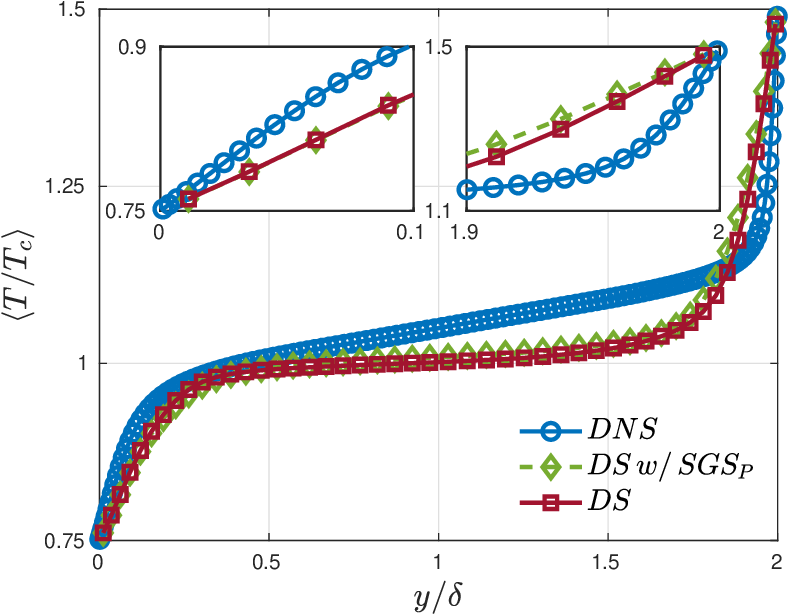}} \\
	\caption{Time-averaged velocity (a) and temperature (b) profiles in the wall-normal direction normalized by channel half-height ($y/\delta$) comparing the SGS stress model alone ($DS \thinspace w/SGS_P$) and coupled with the SGS term of the pressure transport equation ($DS$).} \label{fig:u_T_vs_y_WRLES_SGS}
\end{figure}

To quantify the effect of the SGS terms associated to the pressure transport equation, labelled as $\alpha_4$ and $\alpha_5$, an assessment is performed with these SGS models active and inactive. In this regard, Figure~\ref{fig:u_T_vs_y_WRLES_SGS} depicts the first-order flow statistics between these two cases for the cold and hot walls (coupled with the DS SGS model).
Based on the results, it can be observed that its effect is small, \review{enhancing the velocity profile in the vicinity of the cold wall towards the channel center}, but based on the \textit{a priori} findings, it helps to model the unresolved part of the pressure transport equation.
On the other hand, the ILA model has not been quantified due to the strong effect in the pseudo-boiling region. It is a complex model and it, thus, needs adjustments on the filtering choice (which has been constrained to the adaptive box filter to attain local minimum diminishing property). Nevertheless, based on this framework and the strong thermodynamic gradients when enabling the ILA model, the computation eventually diverge.
While the results were encouraging from an \textit{a priori} standpoint, they were only limited to the inner region. Consequently, the implementation of the model and the \textit{a posteriori} analysis resulted on a too aggressive operation of the model near the walls and in the pseudo-boiling region. In this regard, further research is needed to develop a model suitable for wall-bounded high-pressure transcritical turbulent flow applications.


\section{WMLES SGS stress tensor assessment}\label{sec:Appendix_C}

This appendix compares the performance, in terms of first-order flow statistics, for the different SGS stress tensor models assessed on the various WMLES approaches examined: dynamic, coupled and correlation-based high-pressure models. 

\subsection{Dynamic wall model}

\begin{figure}
	\centering
	\subfloat[]{\includegraphics[width=0.49\linewidth]{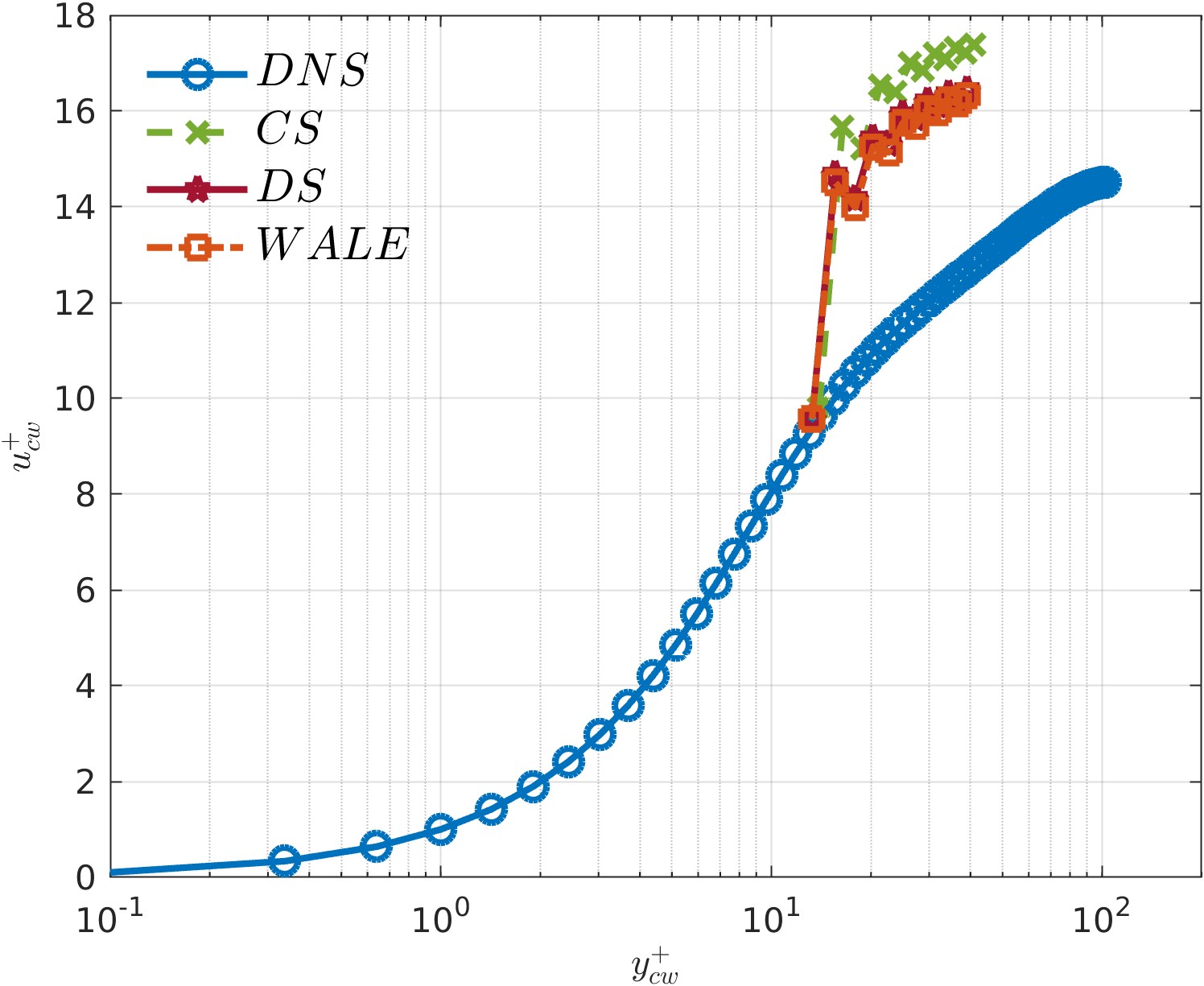}}
    \subfloat[]{\includegraphics[width=0.49\linewidth]{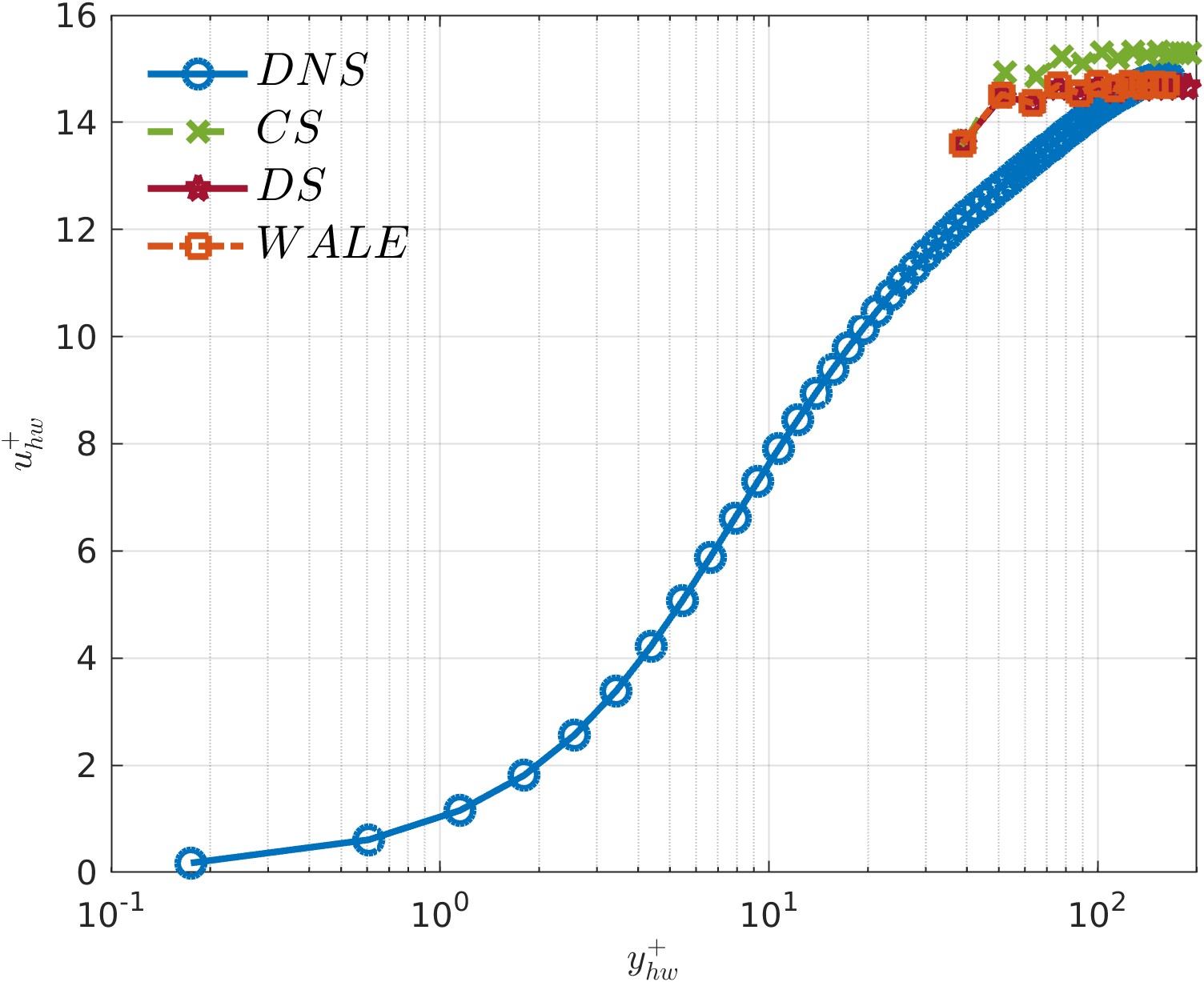}} \\
    \subfloat[]{\includegraphics[width=0.49\linewidth]{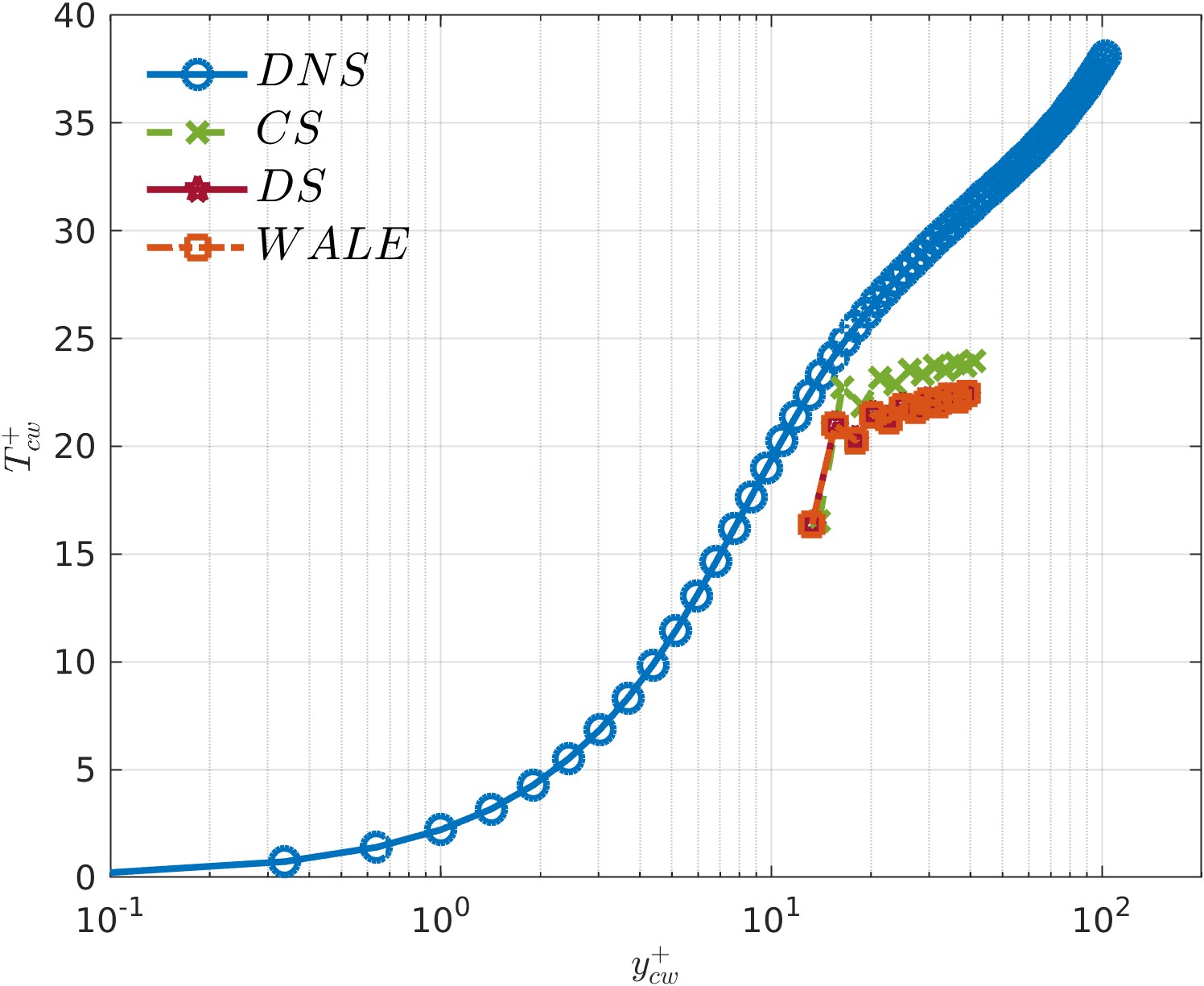}}
    \subfloat[]{\includegraphics[width=0.49\linewidth]{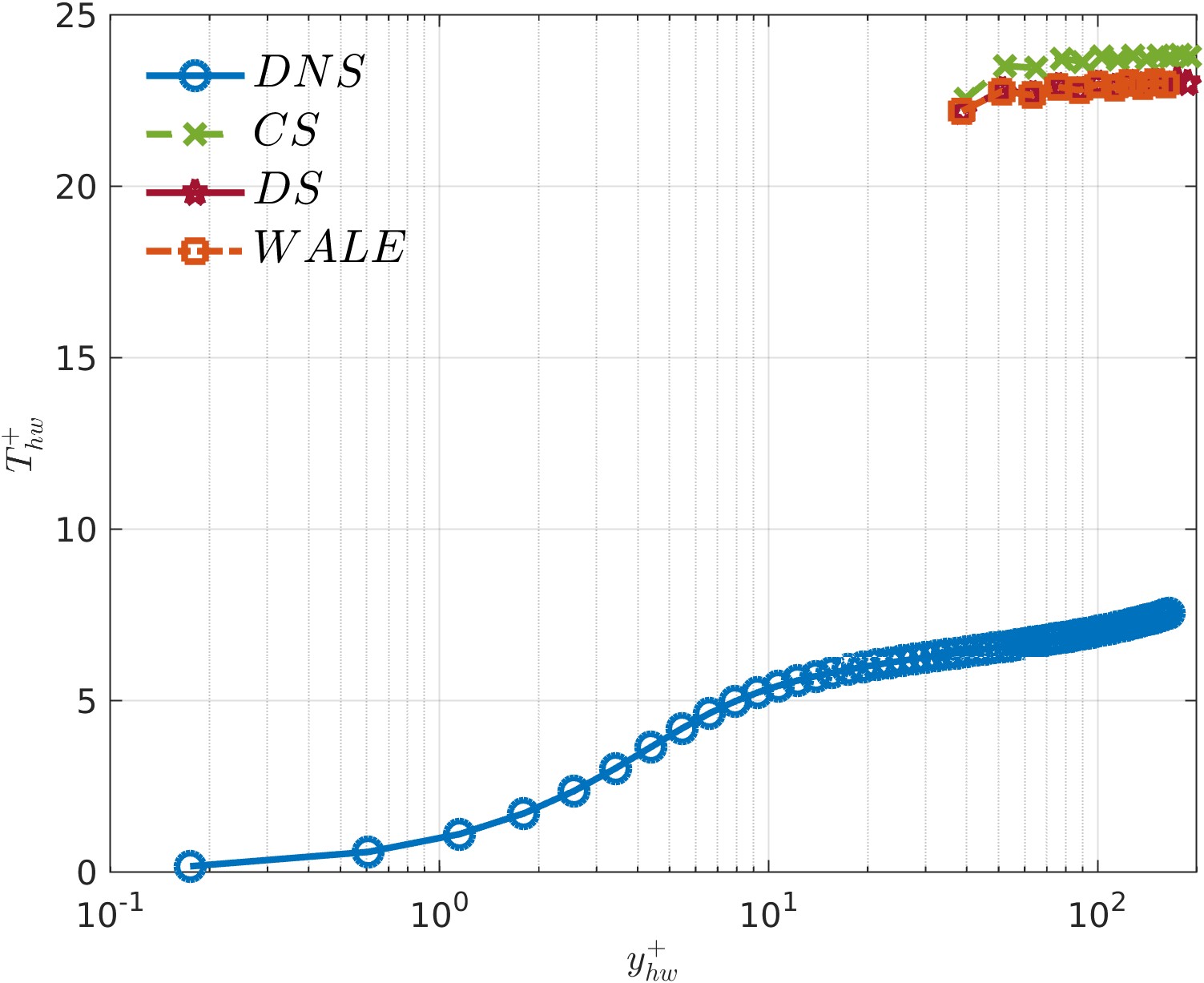}}
	\caption{First-order statistics for velocity (a,b) and temperature (c,d) in wall units comparing WMLES dynamic ``law of the wall'' with CS, DS and WALE models with respect to the DNS dataset for cold (a,c) and hot (b,d) walls.} \label{fig:statistics_WMLES_Dynamic}
\end{figure}

The WMLES-D first-order statistics for velocity and temperature are depicted in Figure~\ref{fig:statistics_WMLES_Dynamic}.

\subsection{Coupled wall model}

\begin{figure}
	\centering
	\subfloat[]{\includegraphics[width=0.485\linewidth]{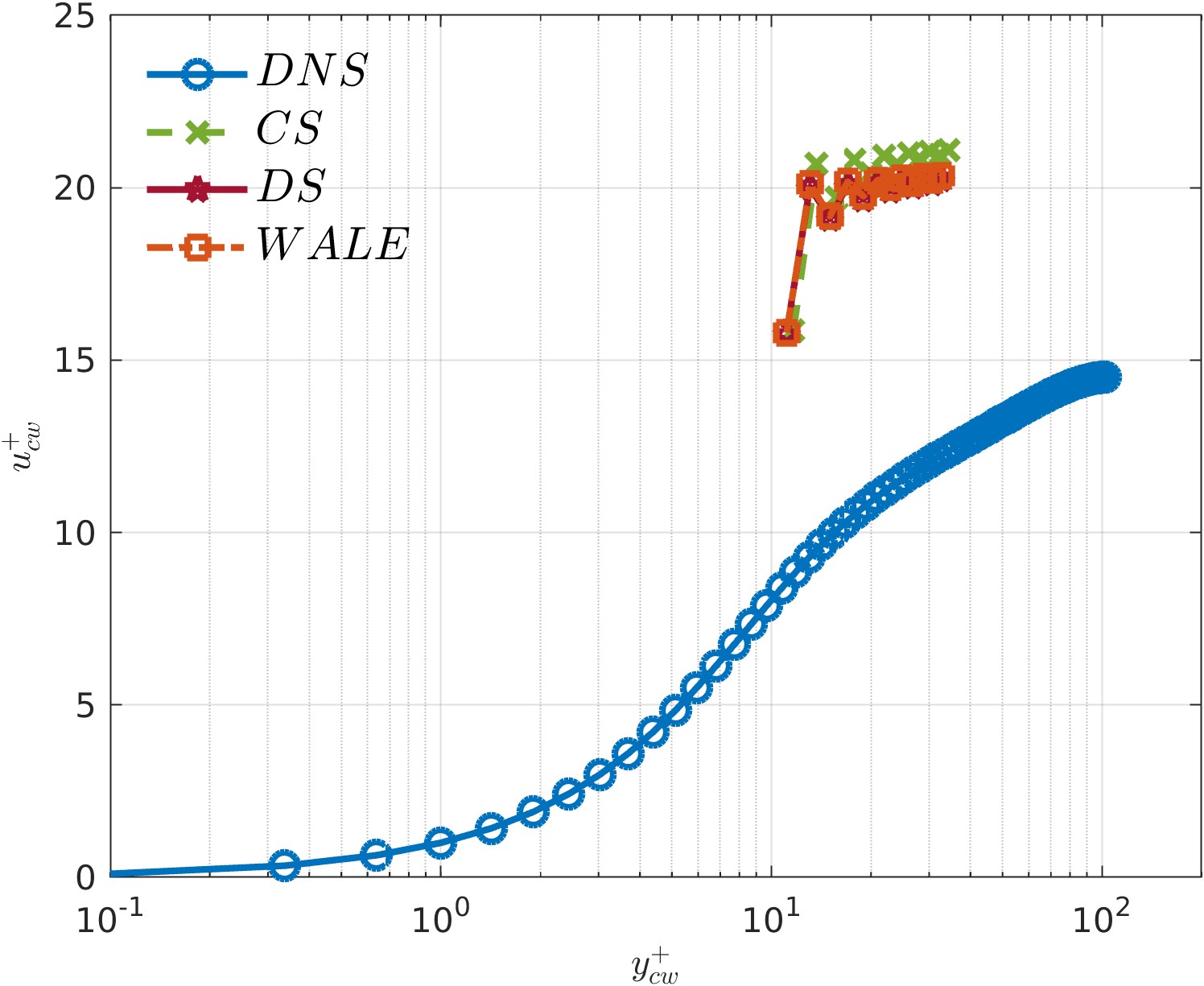}}
    \subfloat[]{\includegraphics[width=0.50\linewidth]{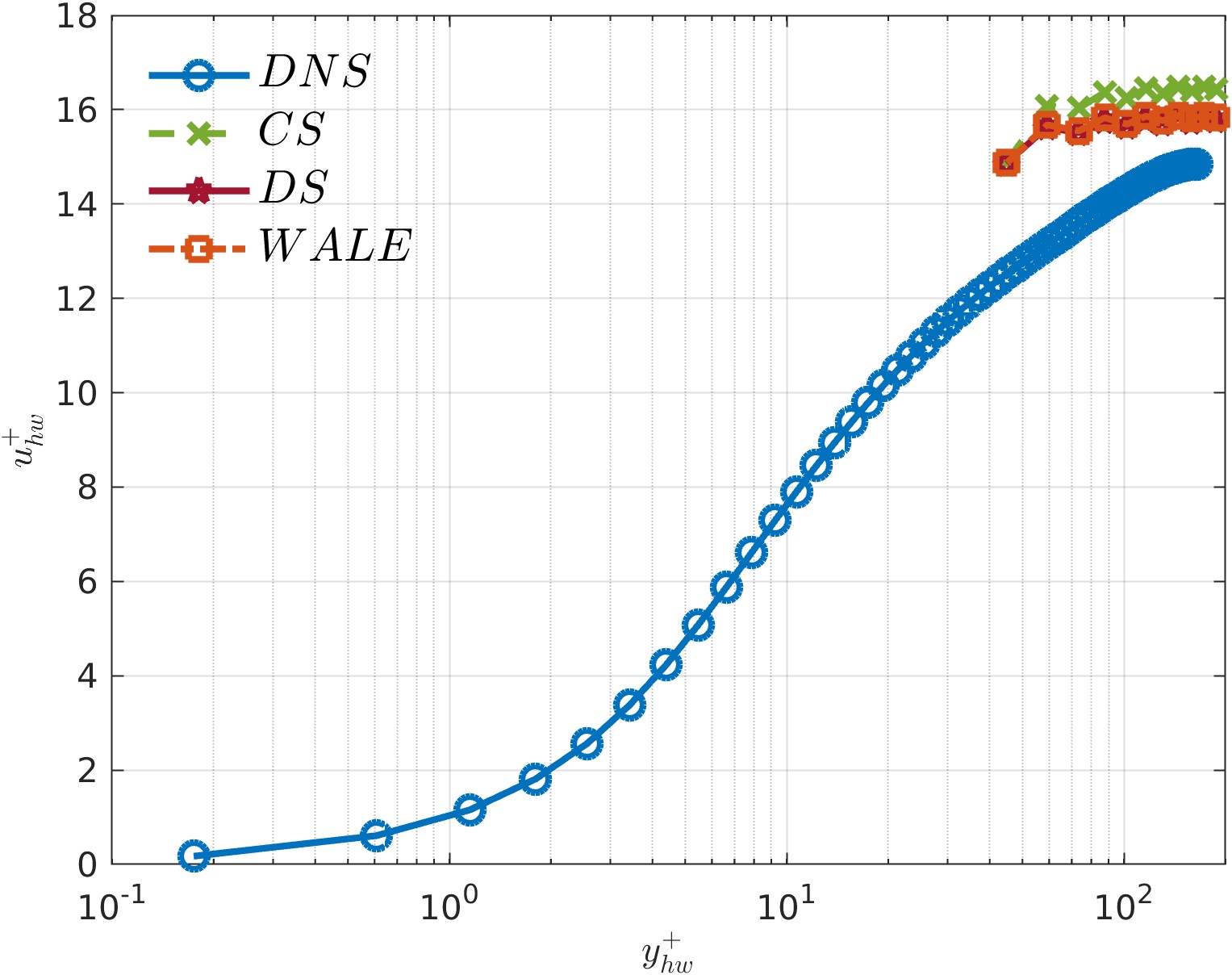}} \\
    \subfloat[]{\includegraphics[width=0.485\linewidth]{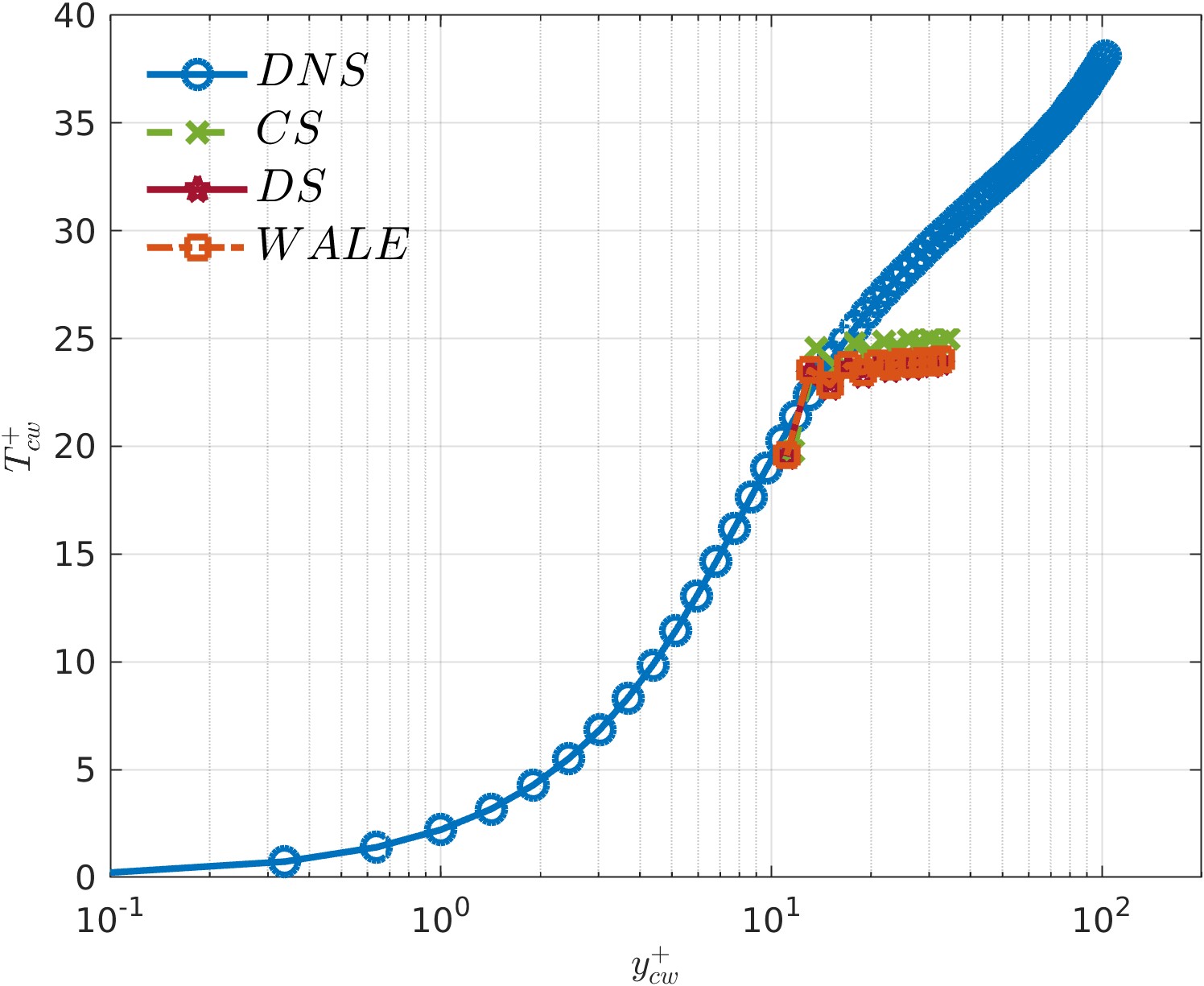}}
    \subfloat[]{\includegraphics[width=0.50\linewidth]{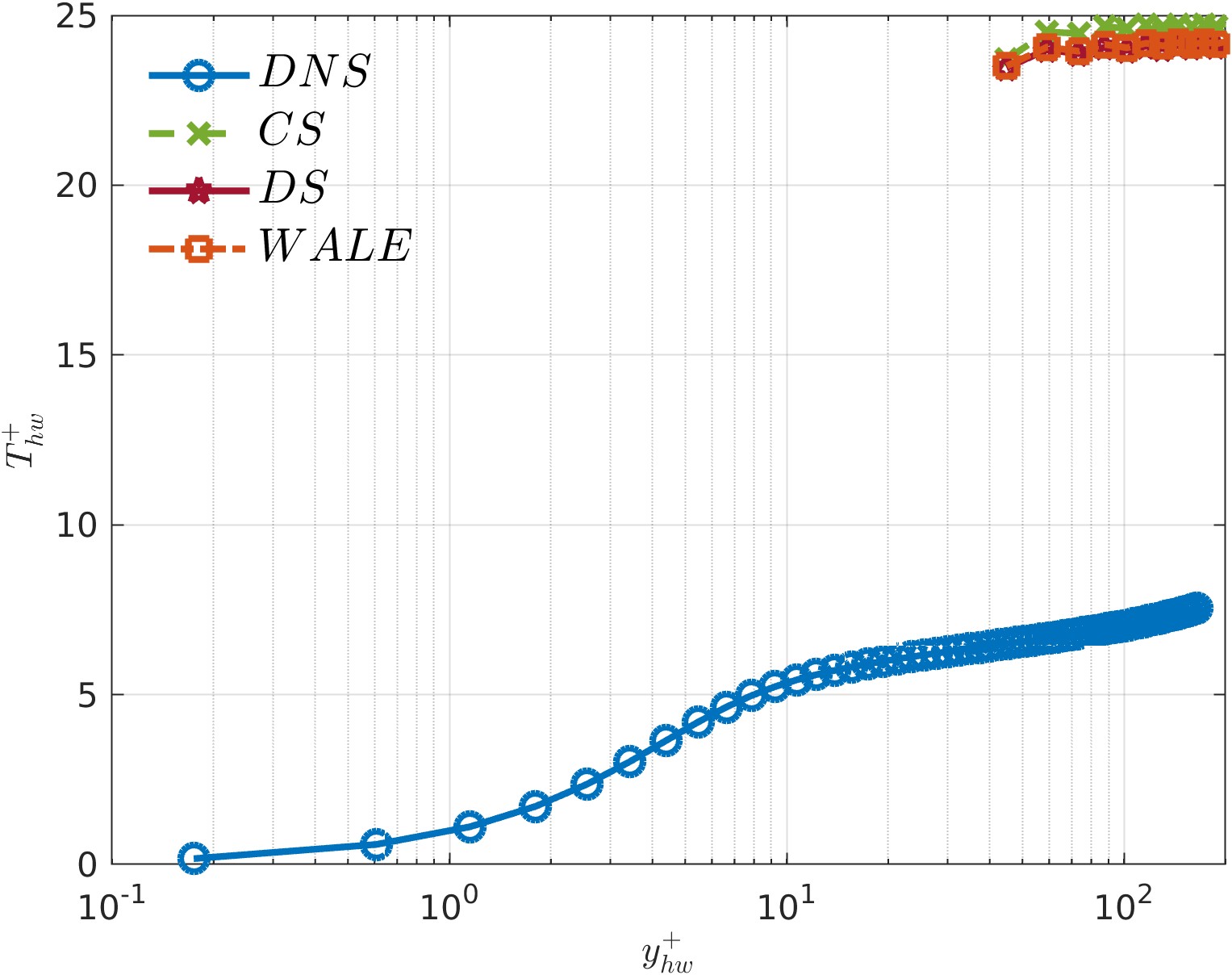}}
	\caption{First order statistics for velocity (a,b) and temperature (c,d) in wall units comparing WMLES coupled law of the wall with CS, DS and WALE models with respect to fully resolved DNS dataset for cold (a,c) and (b,d) hot walls.} \label{fig:statistics_WMLES_Coupled}
\end{figure}

The WMLES-C first-order statistics for velocity and temperature are depicted in Figure~\ref{fig:statistics_WMLES_Coupled}.

\subsection{Correlation-based wall model}

This model (WMLES-HP) is based on a numerical optimization of the first-order flow statistics from DNS with the aim of accurately matching the objective velocity and temperature profiles. The expressions for friction velocity and temperature equations have the following form
\begin{align}
    u^+ & = a_1 \thinspace \textrm{ln} (1 + a_2 y^+) + a_3 \left( 1 - e^{- a_4 y^+} - \frac{y^+}{a_5}e^{- a_6 y^+}\right), \label{eq:u_plus_numeric} \\
    T^+ & = b_1 \thinspace y^+ e^{b_2 y^+/(1 + b_3 y^+)} + \left[ b_4 \thinspace \textrm{ln} (1 + y^+) + b_5 \right] e^{b_2 y^+/(1 + b_3 y^+)}, \label{eq:T_plus_numeric}
\end{align}
whose resulting coefficients from the optimization procedure are gathered in Table~\ref{tab:numeric_optimisation_coefficients}.
\begin{table}

\resizebox{\textwidth}{!}{
\begin{tabular}{ccc|cccccc}
                & & & $i = 1$ & $i = 2$ & $i = 3$ & $i = 4$ & $i = 5$ & $i = 6$\\
    \hline
    \multirow{2}{*}{$a_i$} & & Hot  & $4.80$ & $3.96\cdot 10^{-1}$ & $2.36\cdot 10^{5}$ & $-1.42\cdot 10^{-7}$ & $5.47\cdot 10^{4}$ & $1.74\cdot 10^{5}$  \\
                         &  & Cold & $3.43$ & $1.15$ & $4.44 \cdot 10^{-1}$ & $-1.66\cdot 10^{-2}$ & $1.87\cdot 10^{-1}$ & $3.87\cdot 10^{-1}$ \\
    \hline
    \multirow{2}{*}{$b_i$} & & Hot  & $4.88\cdot 10^{-2}$ & $-2.29 \cdot 10^{-2}$ & $1.63 \cdot 10^{-2}$ & $2.72$ & $-7.44\cdot 10^{-1}$ & $-$  \\
                         &  & Cold & $1.93\cdot 10^{-3}$ & $1.71$ & $3.63 \cdot 10^{-1}$ & $-1.19\cdot 10^{-1}$ & $7.46 \cdot 10^{-1}$ & $-$ \\
    \hline
                                       
  \end{tabular}}

  \caption{Correlation-based optimization coefficients for Eqs.\ref{eq:u_plus_numeric}-\ref{eq:T_plus_numeric} obtained from DNS at the hot and cold walls.} \label{tab:numeric_optimisation_coefficients}
  
\end{table}

Based on this formulation, the WMLES-HP first-order flow statistics are depicted in Figure~\ref{fig:snapshots_spanwise_DNS_WMLES_CS}. The performance among the SGS models is similar to the relative difference observed in Figure~\ref{fig:statistics_WMLES_Dynamic}-\ref{fig:statistics_WMLES_Coupled}; for brevity this overlay comparison is not included in this appendix.


\end{appendices}


\bibliography{References}

\end{document}